\documentclass[a4paper,fleqn,sort&compress]{cas-sc}
\usepackage[numbers,square]{natbib}
\usepackage[center]{caption}
\usepackage{algorithmic}
\usepackage{algorithm}
\usepackage{amsmath}
\usepackage{svg}
\usepackage{wasysym}
\usepackage{soul}
\usepackage{color}
\usepackage{subcaption}
\usepackage{acro}
\DeclareAcronym{BiCGSTAB}{
  short=BiCGSTAB,
  long=biconjugate gradients stabilized,
}
\DeclareAcronym{cfl}{
  short=CFL,
  long=Courant-Friedrichs-Lewy,
}
\DeclareAcronym{cicsam}{
  short=CICSAM,
  long=compressive interface-capturing scheme for arbitrary meshes,
}
\DeclareAcronym{csf}{
  short=CSF,
  long=continuum surface force,
}
\DeclareAcronym{hf}{
  short=HF,
  long=height-function,
}
\DeclareAcronym{mwi}{
  short=MWI,
  long=momentum-weighted interpolation,
}
\DeclareAcronym{rct}{
  short=RCT,
  long=reduced computational time,
}
\DeclareAcronym{thinc}{
  short=THINC/QQ,
  long=tangent of hyperbola interface capturing with quadratic surface representation and Gaussian quadrature,
}
\DeclareAcronym{vof}{
  short=VOF,
  long=volume-of-fluid,
}
\DeclareAcronym{wct}{
  short=WCT,
  long=wall clock time,
}
\usepackage{tikz}
\usetikzlibrary{calc}
\usetikzlibrary{shapes.geometric, arrows}
\usetikzlibrary{arrows.meta}
\tikzset{>={Latex[width=1mm,length=1mm]}}

\tikzstyle{process} = [rectangle, minimum width=2.5cm, minimum height=1cm, text
centered, text width = 3cm, draw=black]
\tikzstyle{decision} = [diamond, minimum width=2.5cm, minimum height=1cm,
aspect=2,inner sep=-0.5ex,text centered, text width = 2.5cm, draw=black]
\tikzstyle{arrow} = [thick,->,>=stealth]
\tikzstyle{line} = [draw, -latex']
\def\tsc#1{\csdef{#1}{\textsc{\lowercase{#1}}\xspace}}
\tsc{WGM}
\tsc{QE}
\tsc{EP}
\tsc{PMS}
\tsc{BEC}
\tsc{DE}
\begin{document}
\let\WriteBookmarks\relax
\def\floatpagepagefraction{1}
\def\textpagefraction{.001}
% \shorttitle{A fully-coupled algorithm for interfacial flows with implicit surface tension}
\shorttitle{A fully-coupled algorithm with implicit surface tension treatment for interfacial flows}
\shortauthors{R. Janodet et~al.}

% \title [mode = title]{A fully-coupled algorithm for interfacial flows with implicit surface tension}
\title [mode = title]{A fully-coupled algorithm with implicit surface tension treatment for interfacial flows with large density ratios}

\author[1]{Romain Janodet}[orcid=0000-0001-5459-7891]
\author[1]{Berend {van Wachem}}[orcid=0000-0002-5399-4075]%\fnref{fn3}}
\author[2]{Fabian Denner}[orcid=0000-0001-5812-061X]
\cormark[1]
\ead{fabian.denner@polymtl.ca}

\address[1]{Chair of Mechanical Process Engineering, Otto-von-Guericke-Universität Magdeburg, Universitätsplatz 2, 39106 Magdeburg, Germany}

\address[2]{Department of Mechanical Engineering, Polytechnique Montréal, 2500 chemin de Polytechnique, Montréal, H3T 1J4, Québec, Canada}

\cortext[cor1]{Corresponding author}

%------------------------------------------------------------
\begin{abstract}
The stability of most surface-tension-driven interfacial flow simulations is governed by the capillary time-step constraint. This concerns particularly small-scale flows and, more generally, highly-resolved liquid-gas simulations with moderate inertia. To date, the majority of interfacial-flow simulations are performed using an explicit surface-tension treatment, which restrains the performance of such simulations. Recently, an implicit treatment of surface tension able to breach the capillary time-step constraint using the volume-of-fluid (VOF) method was proposed, based on a fully-coupled pressure-based finite-volume algorithm. To this end, the interface-advection equation is incorporated implicitly into the linear flow solver, resulting in a tight coupling between all implicit solution variables (colour function, pressure, velocity). However, this algorithm is limited to uniform density and viscosity fields. Here, we present a fully-coupled algorithm for interfacial flows with implicit surface tension applicable to interfacial flows with large density and viscosity ratios. This is achieved by solving the continuity and momentum equations in conservative form, whereby the density is treated implicitly with respect to the colour function, and the advection term of the interface-advection equation is discretised using the THINC/QQ algebraic VOF scheme, yielding a consistent discretisation of the advective terms. This new algorithm is tested by considering representative surface-tension-dominated interfacial flows, including the Laplace equilibrium of a stationary droplet and the three-dimensional Rayleigh-Plateau instability of a liquid filament. The presented results demonstrate that interfacial flows with large density and viscosity ratios can be simulated and energy conservation is ensured, even with a time step larger than the capillary time-step constraint, provided that other time-step restrictions are satisfied.
\end{abstract}

\begin{keywords}
Multiphase flows \sep Surface tension \sep Capillary time-step constraint \sep Coupled algorithm \sep Volume-of-fluid method \sep THINC method %\sep Finite volume
\end{keywords}

\maketitle

%\begin{small}
    % \printacronyms
%\end{small}

%------------------------------------------------------------
\section{Introduction}
\label{section:intro}

The maximum stable numerical time step $\Delta t$ which is possible for the simulation of surface-tension-driven interfacial flow simulations is determined by various stability criteria, each one being driven by a physical phenomenon and associated with a scaling with respect to the mesh spacing $\Delta x$. For an incompressible flow, the time-step constraint associated with inertia is determined from the well-known \ac{cfl} condition~\citep{Courant1928},
    \begin{equation}
        \Delta t_\text{CFL} = \frac{\mathrm{CFL}\,\Delta x}{|\mathbf{u}|_\text{max}} \propto \Delta x,
        \label{eq:CFL_theory}
    \end{equation}
where $\mathbf{u}$ is the fluid velocity vector and $\mathrm{CFL}$ is the CFL number, the time-step constraint associated with diffusion (mechanical, thermal, chemical) is determined by the Fourier number $\mathrm{Fo}$ \citep{Patankar1980}, 
    \begin{equation}
        \Delta t_\text{Fo} = \frac{\mathrm{Fo}\,\Delta x^2}{a} \propto \Delta x^2,
        \label{eq:Fo_theory}
    \end{equation}
where $a$ is the relevant diffusivity, and the time-step constraint associated with surface tension is determined, as originally formulated by~\citet{Brackbill1992}, by the so-called capillary time-step constraint,
\begin{equation}
        \Delta t_\sigma = \sqrt{\frac{\rho_\mathrm{A} + \rho_\mathrm{B}}{4\pi\sigma}\Delta x^3} \propto \Delta x^{3/2},
        \label{eq:dt_sigma_theory_brackbill}
    \end{equation}
where $\rho_\mathrm{A}$ and $\rho_\mathrm{B}$ are the densities of the two interacting fluids A and B, respectively, and $\sigma$ is the surface tension coefficient. 
This criterion is in fact a CFL condition associated with the phase velocity $c_\sigma$ of the smallest unambiguously resolved capillary waves: $\Delta t_\sigma \propto \Delta x/c_\sigma$.
\citet{Denner2015} revisited the capillary time-step constraint using both numerical and signal-processing approaches, and found a slightly less restrictive formulation,  
\begin{equation}
    \Delta t_\sigma = \sqrt{\frac{\rho_\mathrm{A} + \rho_\mathrm{B}}{2\pi\sigma}\Delta x^3} \propto \Delta x^{3/2},
    \label{eq:dt_sigma_theory_Denner}
\end{equation}
compared to the original formulation given in Eq.~\eqref{eq:dt_sigma_theory_brackbill}. Although the diffusion time-step constraint shows the strongest scaling with respect to the mesh spacing, contemporary numerical algorithms for incompressible flows solve diffusion terms implicitly, thereby eliminating the diffusion time-step constraint. Consequently, the capillary time-step constraint is generally the dominant time-step constraint for interfacial flows at small scales~\citep{Galusinski2008,Denner2015,Popinet2018}.

Since Brackbill's hypothesis that an implicit discretisation of surface tension should allow breaching or removing the capillary time-step constraint, the community has worked towards this goal. The first studies that presented numerical methods that are able to breach the capillary time-step constraint include the boundary integral method with an implicit surface tension treatment for two-dimensional irrotational incompressible ﬂows of~\citet{Hou1994a} and the works of~\citet{Bansch2001},~\citet{Hysing2006} and~\citet{Raessi2009} on a semi-implicit surface tension formulation, which includes the interface location at the new time instance implicitly. With the latter, stable results for time steps larger than the capillary time-step constraint were obtained, but the solution is not stable for arbitrarily large time steps. As demonstrated by~\citet{Denner2017}, the method of \citet{Hysing2006} and~\citet{Raessi2009} is equivalent to the addition of a surface viscosity that attenuates fast capillary waves. \citet{Sussman2009} proposed a method that works in a similar manner as the scheme of~\citet{Hysing2006}, by adding surface damping acting as a low-pass filter~\citep{Popinet2018}.
{\color{black} \citet{Zheng2015} proposed an implicit Lagrangian method that is not bound by the capillary time-step constraint and which, in contrast to the majority of methods addressing the capillary time-step constraint that express surface tension as a volumetric source term \citep{Bansch2001,Hysing2006,Raessi2009,Denner2022b}, retains a sharp representation of the interface.}
% , and \citet{Zheng2015} proposed an implicit Lagrangian method that is not bound by the capillary time-step constraint.

Recently, \citet{Denner2022b} proposed a fully-coupled algorithm with an implicit treatment of surface tension that is able to breach, without numerical artifacts, the capillary time-step constraint. This was achieved by employing a Newton linearisation of the surface-tension source term with the \ac{csf} approach~\citep{Brackbill1992}, and expressing the implicit interface curvature and colour-function gradients with respect to the implicit colour function. This methodology provides a tight coupling between all implicit solution variables (colour function, pressure, velocity), by applying an algebraic volume-of-fluid (VOF) method for the discretisation of the interface-advection equation and solving the discretised interface-advection equation together with the discretised \textcolor{black}{continuity} and momentum equations simultaneously as an implicitly coupled system of equations. However, although the study of \citet{Denner2022b}  demonstrates that the capillary time-step constraint can be breached by treating surface tension implicitly, their algorithm is only able to simulate interfacial flows with unit density and viscosity ratios, and the interface advection is discretised by the \ac{cicsam}~\citep{Ubbink1999}, which is known to require very small CFL numbers to retain a sharp interface~\citep{Gopala2008}.

In this study, a fully-coupled algorithm for interfacial flows with implicit surface tension is presented that is able to simulate interfacial flows with realistic density and viscosity ratios, and breach the capillary time-step constraint. In order to enable this algorithm for flows with realistic density and viscosity ratios, the continuity and momentum equations are discretised in conservative form, with a consistent discretisation of the advective terms in all governing equations and where density is treated implicitly with respect to the colour function in the transient terms. Moreover, in order to mitigate the CFL time-step constraint associated with the interface transport, the THINC/QQ scheme \citep{Xie2017,Chen2022b} is applied to discretise the advection term of the interface-advection equation, as it generally can handle larger time steps than the CICSAM scheme, while retaining a sharp fluid interface. To test and scrutinize this new algorithm, two- and three-dimensional interfacial flows with surface tension and large density ratios are considered, ranging from the two-dimensional Laplace equilibrium to a three-dimensional Rayleigh-Plateau instability.

The governing equations are introduced in Section \ref{section:maths} and the proposed numerical framework is presented in Section \ref{section:numerics}. Section \ref{section:upper_limit} briefly revisits the less stringent time-step constraint associated with surface tension encountered with the proposed algorithm. The results of four representative test cases are presented and discussed in Section \ref{section:validation} and the article is concluded in Section \ref{section:conclusion}.

\section{Governing equations}
\label{section:maths}
The governing equations describing immiscible interfacial flows with surface tension in the one-fluid formulation are the continuity equation,
\begin{equation}
    \frac{\partial\rho}{\partial t} + \boldsymbol{\nabla}\cdot (\rho\mathbf{u}) = 0,
    \label{eq:NS_mass_cons}
\end{equation}
where $\mathbf{u}$ is the velocity vector and $\rho$ is the density, and the momentum equations,
\begin{equation}
    \frac{\partial(\rho\mathbf{u})}{\partial t} + \boldsymbol{\nabla}\cdot (\rho\mathbf{u}\otimes \mathbf{u}) = - \boldsymbol{\nabla} p + \boldsymbol{\nabla}\cdot \boldsymbol{\tau} + \mathbf{S}_\sigma,
    \label{eq:NS_momentum_cons}
\end{equation}
where $p$ is the pressure, $\boldsymbol{\tau}$ is the viscous stress tensor, and $\mathbf{S}_\sigma$ is the volumetric source term representing surface tension. Since the aim of this work is to simulate incompressible interfacial flows with large density ratios, the conservative forms of both the continuity and momentum equations are considered~\citep{Rudman1998}.
For an isotropic Newtonian fluid, the viscous stress tensor $\boldsymbol{\tau}$ is defined as
\begin{equation}
    \boldsymbol{\tau} = \mu \left(\boldsymbol{\nabla}\mathbf{u} + \boldsymbol{\nabla}\mathbf{u}^T\right),
    \label{eq:NS_tau}
\end{equation}
with $\mu$ the dynamic viscosity. The surface-tension source term $\mathbf{S}_\sigma$ acts at the fluid interface between two immiscible liquids and is defined as~\citep{Tryggvason2011a}
\begin{equation}
    \mathbf{S}_\sigma = \sigma \kappa \mathbf{n}_\mathrm{S} \delta_\mathrm{S},
    \label{eq:surface_tension_source_maths}
\end{equation}
where $\sigma$ is the constant surface-tension coefficient, $\kappa$ is the interface curvature, $\mathbf{n}_\mathrm{S}$ is the interface normal vector, and $\delta_\mathrm{S}$ is the Dirac delta function associated with the interface.

A volume-of-fluid (VOF) method~\citep{Hirt1981} is employed to model the transport and interaction of two immiscible fluids. The indicator function $H$, used here to distinguish two immiscible fluid phases, is defined at position $\mathbf{x}$ as 
\begin{equation}
    H (\mathbf{x}) = 
    \begin{cases}
    1\,\,\mathrm{if}\,\,\mathbf{x}\,\,\text{in in fluid A} \\ 
    0\,\,\mathrm{if}\,\,\mathbf{x}\,\,\text{is in fluid B}
    \end{cases}
    \label{eq:phase_indicator_def}
\end{equation}
where fluids A and B may represent any immiscible combination of fluids, and advected with the flow,
\begin{equation}
    \frac{DH}{Dt} \textcolor{black}{= \frac{\partial H}{\partial t} + \mathbf{u}\cdot \boldsymbol{\nabla}H = \frac{\partial H}{\partial t} + \boldsymbol{\nabla}\cdot (H \mathbf{u}) - H \boldsymbol{\nabla}\cdot \mathbf{u}} = 0.
    \label{eq:indicator_advection}
\end{equation}
\textcolor{black}{Since the sharp Heaviside function $H$ cannot be convected numerically on a grid, the colour function $\psi$, defined as the control-volume average of $H$ as}
\begin{equation}
    \textcolor{black}{\psi = \frac{1}{V} \int_{V} H \, \mathrm{d}V},
    \label{eq:colour_fct_def_math}
\end{equation}
\textcolor{black}{is convected instead. By integrating Eq.~(\ref{eq:indicator_advection}) over the control volume $V$ with boundary $\partial V$, Eq.~(\ref{eq:indicator_advection}) becomes the advection equation for the colour function}
\begin{equation}
    \textcolor{black}{\int_V \frac{\partial \psi}{\partial t} \, \mathrm{d}V + \int_{\partial V} \left(H \mathbf{u}\right) \cdot \mathbf{d}\boldsymbol{A} - \int_V \left(H \boldsymbol{\nabla}\cdot \mathbf{u}\right) \, \mathrm{d}V = 0.}
    \label{eq:colour_advection}
\end{equation}
% Because, in practice, immiscible interfacial flows with surface tension are typically gas-liquid flows, we assume in the following that fluid A is a liquid and fluid B is a gas. 
The fluid properties are defined over the entire computational domain based on the individual properties of the two fluids, denoted with subscripts $\mathrm{A}$ and $\mathrm{B}$, respectively, and the \textcolor{black}{colour function $\psi$} as
\begin{align}
    \rho(\mathbf{x})  &= \rho_\mathrm{B} + \textcolor{black}{\psi(\mathbf{x})} \, (\rho_\mathrm{A} - \rho_\mathrm{B}) \\
    \mu(\mathbf{x})  &= \mu_\mathrm{B} + \textcolor{black}{\psi(\mathbf{x})} \, (\mu_\mathrm{A} -\mu_\mathrm{B}).
    \label{eq:fluid_prop_everywhere}
\end{align}
\textcolor{black}{such that $\rho$ and $\mu$ are consistent with the interface position.}

\section{Coupled numerical framework}
\label{section:numerics}

The proposed fully-coupled algorithm for interfacial flows solves the discretised and linearised governing equations presented in Section \ref{section:maths} simultaneously and implicitly coupled in a single system of linear equations, $\boldsymbol{\mathcal{A}} \cdot \boldsymbol{\phi} = \mathbf{b}$. To this end, the governing equations are discretised using a second-order finite-volume method with a collocated variable arrangement \citep{Denner2020}, whereby the interface advection is discretised using the THINC/QQ scheme \citep{Xie2017,Chen2022b} and the volume flux through the mesh faces is computed using a momentum-weighted interpolation (MWI)~\citep{Bartholomew2018} and treated implicitly. 

\subsection{Implicit surface-tension source term}
\label{subsection:discretisation_surf_tens}

With the colour function $\psi$, the volumetric source term representing surface tension, Eq.~\eqref{eq:surface_tension_source_maths}, is described by the CSF model~\citep{Brackbill1992}, defined for cell $P$ as
\begin{equation}
    \mathbf{S}_{\sigma,P} = \sigma \, \kappa_P \, \boldsymbol{\nabla} \psi_P,
    \label{eq:csf}
\end{equation}
where the surface tension coefficient $\sigma$ is assumed to be constant.
The gradient of the colour function $\psi$ is discretised using the Gauss theorem, given for a cell $P$ as
\begin{equation}
     \boldsymbol{\nabla} \psi_P \approx \frac{1}{V_P} \sum_f \overline{\psi}_f \mathbf{n}_f A_f ,
    \label{eq:colourgrad}
 \end{equation}
where $\overline\square$ denotes a linear interpolation, $f$ denotes the faces adjacent to cell $P$, and $\mathbf{n}_f$ and $A_f$ are the outward pointing unit normal vector and the area of face $f$, respectively. In order to treat the source term $\mathbf{S}_\sigma$ implicitly, a Newton linearisation is applied to the colour function gradients and the interface curvature, such that \citep{Denner2022b}
\begin{equation}
        \mathbf{S}_{\sigma,P}^{(n+1)} \approx \frac{\sigma}{V_P} \left(\kappa_P^{(n)} \sum_f \overline{\psi}_f^{(n+1)} \mathbf{n}_{f} A_f + \kappa_P^{(n+1)}  \sum_f \overline{\psi}_f^{(n)} \mathbf{n}_{f} A_f -  \kappa_P^{(n)}  \sum_f \overline{\psi}_f^{(n)} \mathbf{n}_{f} A_f  \right) 
    \label{eq:csf_imp},
\end{equation}
where $n$ is the nonlinear iteration counter, with superscript $(n)$ denoting deferred quantities and superscript $(n+1)$ denoting an implicitly solved quantity. 

Following \citet{Denner2022b}, an implicit formulation of the \ac{hf} method~\citep{Evrard2020} is applied to compute the interface curvature. Using the HF method and assuming the $z$-component of the interface normal vector is dominant, the interface curvature is given as
\begin{equation}
    \kappa = \frac{-h_{xx}\left(1+h_y^2\right) - h_{yy}\left(1+h_x^2\right)+2h_xh_yh_{xy}}{\left(h_x^2+h_y^2+1\right)^{3/2}} = \frac{\mathcal{M}}{\mathcal{D}^{3/2}}, \label{eq:hf}
\end{equation}
where $h_{\{x,y,xx,yy,xy\}}$ denotes the first and second partial derivatives of the liquid heights computed along the $z$-direction. By permuting the indices, the curvature can be computed in the same way for cases in which either the  $x$- or $y$-component of the interface normal vector is the dominant component. 
Applying a Newton linearisation to Eq.~\eqref{eq:hf}, the implicit interface curvature is given as
\begin{equation}
\kappa^{(n+1)}_P = \kappa^{(n)}_P + \frac{1}{\mathcal{D}_P^{(n)}} \sum_{N\in\mathcal{S}(P)} \left[\left(\psi^{(n+1)}_N-\psi^{(n)}_N\right) \left(\left(\frac{\partial \mathcal{M}_P}{\partial \psi_N}\right)^{(n)} - \frac{3}{2} \kappa^{(n)}_P \sqrt{\mathcal{D}_P^{(n)}} \left(\frac{\partial \mathcal{D}_P}{\partial \psi_N}\right)^{(n)}\right) \right],
\label{eq:Klinear}
\end{equation}
which is applied on a stencil with $3 \times 3 \times N_h$ cells, where $N_h$ is an odd number between $5$ and $9$.
\textcolor{black}{A detailed derivation of this implicit height-function method can be found in the work of \citet{Denner2022b}.}

\subsection{Discretisation of convective terms}
\label{subsection:convective_terms_discretisation}

In the proposed finite-volume framework, all advective terms take the form
\begin{equation}
    \sum_f \mathcal{F}_f^{(\phi)} =\sum_f \widetilde{\phi}_f^{(n)} F_f^{(n+1)},
    \label{eq:convective_term_generic}
\end{equation}
where $\widetilde{\phi}_f^{(n)}$ is the deferred face-based transported quantity $\phi$ ($\psi$, $\rho$, $\rho\mathbf{u}$) at nonlinear iteration $n$, and $F_f^{(n+1)}$ is the implicitly solved volume flux, calculated using the momentum-weighted interpolation (MWI) presented in Section~\ref{subsubsection:implicit_mwi}. Thus, there are three advective fluxes $\mathcal{F}_f^{(\phi)}$ to compute, which should be all consistent~\citep{Fuster2018,  Arrufat2021}:
\begin{itemize}
    \item Colour function: $\mathcal{F}_f^{(\psi)} = \widetilde{\psi}_f^{(n)} F_f^{(n+1)}$, with $\widetilde{\psi}_f^{(n)}$ constructed using the THINC/QQ scheme~\citep{Xie2017,  Chen2022b}, described in Section~\ref{subsubsection:THINC_QQ}.
    \item Density: $\mathcal{F}_f^{(\rho)} = \widetilde{\rho}_f^{(n)} F_f^{(n+1)} = \rho_\text{A} \mathcal{F}_f^{(\psi)} + \rho_\text{B} \mathcal{F}_f^{(1-\psi)}$, which depends on the flux of the colour function for consistency, $\widetilde{\rho}_f^{(n)} = \widetilde{\rho}_f^{(n)}(\widetilde{\psi}_f^{(n)})$, as further discussed in Section~\ref{subsubsection:flux_density_momentum}.
    \item Momentum: $\mathcal{F}_f^{(\rho\mathbf{u})} = \widetilde{(\rho\mathbf{u})}_f^{(n)} F_f^{(n+1)} = \widetilde{\rho}_f^{(n)} \widetilde{\mathbf{u}}_f^{(n)}  F_f^{(n+1)} = \mathcal{F}_f^{(\rho)}\widetilde{\mathbf{u}}_f^{(n)}$, which depends on the density flux for consistency, as described in Section~\ref{subsubsection:flux_density_momentum}.
\end{itemize}
\textcolor{black}{It should be noted that, contrary to \citet{Denner2022b} where we applied a Newton linearisation to the convective terms, here we apply a Picard linearisation with the volume flux $F_f$ being treated implicitly. This Picard linearisation is chosen because the fluxes of the THINC/QQ scheme cannot be written implicitly and, to ensure the consistent transport of mass, momentum, and the fluid interface, this Picard linearisation is applied to all convective terms.}

\subsubsection{Calculation of the volume flux $F_f^{(n+1)}$}
\label{subsubsection:implicit_mwi}

The volume flux $F_f = \vartheta_f A_f$ through cell face $f$ is defined based on an advecting velocity $\vartheta_f=\mathbf{u}_f \cdot \mathbf{n}_f$ that is discretised using a momentum-weighted interpolation (MWI) \citep{Rhie1983, Bartholomew2018}. As previously demonstrated \citep{Denner2018c,Denner2020,Denner2022b}, in the class of fully-coupled algorithms considered in this study, the volume flux can play a major role in implicitly coupling the governing equations. 

Contrary to previously presented fully-coupled algorithms for interfacial flows \citep{Darwish2007,Denner2014a,Darwish2015,Denner2015,Denner2018b,Denner2022b} that include (partially) deferred  volume fluxes, the proposed algorithm only contains fully implicit velocity and pressure terms, and the surface tension terms are made implicit using a Newton linearisation. Including the linearised surface-tension source term as presented in Section \ref{subsection:discretisation_surf_tens}, the implicit advecting velocity is defined by the MWI as \citep{Denner2022b}
\begin{equation}
    \begin{split}
        \vartheta_f^{(n+1)} = \overline{\mathbf{u}}_{f}^{(n+1)} \cdot \mathbf{n}_{f} &-~\hat{d}_f \left[\frac{p_Q^{(n+1)}-p_P^{(n+1)}}{\Delta x} - \frac{\breve{\rho}_f^{(n)}}{2} \left( \frac{\boldsymbol{\nabla} p_P^{(n+1)}}{\rho_P^{(n)}} +  \frac{\boldsymbol{\nabla} p_Q^{(n+1)}}{\rho_Q^{(n)}} \right) \cdot \mathbf{n}_{f} \right] \\ 
        &+~\hat{d}_f \sigma \left[\overline{\kappa}_f^{(n)} \frac{\psi_Q^{(n+1)}-\psi_P^{(n+1)}}{\Delta x} - \frac{\breve{\rho}_f^{(n)}}{2} \left(\kappa_P^{(n)} \frac{\boldsymbol{\nabla} \psi_P^{(n+1)}}{\rho_P^{(n)}} + \kappa_Q^{(n)}\frac{\boldsymbol{\nabla} \psi_Q^{(n+1)}}{\rho_Q^{(n)}} \right)  \cdot \mathbf{n}_{f} \right]\\ 
        &+~\hat{d}_f \sigma \left[\overline{\kappa}_f^{(n+1)} \frac{\psi_Q^{(n)}-\psi_P^{(n)}}{\Delta x} - \frac{\breve{\rho}_f^{(n)}}{2} \left(\kappa_P^{(n+1)} \frac{\boldsymbol{\nabla} \psi_P^{(n)}}{\rho_P^{(n)}} + \kappa_Q^{(n+1)} \frac{\boldsymbol{\nabla} \psi_Q^{(n)}}{\rho_Q^{(n)}} \right)  \cdot \mathbf{n}_{f} \right]\\ 
        &-~\hat{d}_f \sigma \left[\overline{\kappa}_f^{(n)} \frac{\psi_Q^{(n)}-\psi_P^{(n)}}{\Delta x} - \frac{\breve{\rho}_f^{(n)}}{2} \left(\kappa_P^{(n)} \frac{\boldsymbol{\nabla} \psi_P^{(n)}}{\rho_P^{(n)}} + \kappa_Q^{(n)} \frac{\boldsymbol{\nabla} \psi_Q^{(n)}}{\rho_Q^{(n)}} \right)  \cdot \mathbf{n}_{f} \right] \\ 
        &+~\hat{d}_f \frac{\breve{\rho}_f^{(t-\Delta t)}}{\Delta t} \left( \vartheta_f^{(t-\Delta t)} - \overline{\mathbf{u}}_{f}^{(t-\Delta t)} \cdot \mathbf{n}_{f} \right)
    \end{split}
    \label{eq:mwi_implicit}
\end{equation}
where coefficient $\hat{d}_f$ represents the weighting factor of the MWI correction, as derived by \citet{Bartholomew2018}, and the face density $\breve{\rho}_f$ is defined by a harmonic average. 
\begin{equation}
    \frac{1}{\breve{\rho}_f^{(n)}} = \frac{1}{2\rho_P^{(n)}}+\frac{1}{2\rho_Q^{(n)}}.
    \label{eq:hat_face_density_mwi}
\end{equation}

{\color{black} The primary motivation for including the pressure in the computation of the fluxes is to couple pressure and velocity for the employed collocated variable arrangement. To this end, the discretised pressure terms together constitute a low-pass filter on the pressure field that prevents pressure-velocity decoupling \citep{Bartholomew2018},
\begin{equation}
   \boldsymbol{\nabla} p_f - \overline{\boldsymbol{\nabla} p}_f =  \frac{p_Q-p_P}{\Delta x} - \frac{\boldsymbol{\nabla} p_P + \boldsymbol{\nabla} p_Q}{2} \propto \left. \frac{\partial^3 p}{\partial x^3} \right|_f,
   \label{eq:filter_pressure_mwi}
\end{equation}
where the overbar denotes a linear interpolation of cell-centred values. In order to ensure a balance between surface tension and pressure, the source term representing surface tension must be incorporated in the momentum-weighted interpolation in the same fashion as the pressure \citep{Denner2014a,Bartholomew2018},
\begin{equation}
   \sigma \kappa_f \boldsymbol{\nabla} \psi_f - \sigma \overline{\kappa}_f \overline{\boldsymbol{\nabla} \psi}_f =  \sigma \overline{\kappa}_f \frac{\psi_Q-\psi_P}{\Delta x} - \sigma \, \frac{\kappa_P \boldsymbol{\nabla} \psi_P + \kappa_Q \boldsymbol{\nabla} \psi_Q}{2}.
   \label{eq:filter_surf_tens_mwi}
\end{equation}
Since we wish to treat both the interface curvature and the colour function gradient implicitly with respect to the colour function, we further linearise the surface tension terms with a Newton linearisation,
\begin{equation}
   \sigma \kappa^{(n+1)} \boldsymbol{\nabla} \psi^{(n+1)} \approx   \sigma \kappa^{(n+1)} \boldsymbol{\nabla} \psi^{(n)} +   \sigma \kappa^{(n)} \boldsymbol{\nabla} \psi^{(n+1)} -   \sigma \kappa^{(n)} \boldsymbol{\nabla} \psi^{(n)},
   \label{eq:newton_linearisation_surf_tens}
\end{equation}
 similar to Eq.~\eqref{eq:csf_imp}. The density weighting applied to both the pressure and surface tension terms in Eq.~\eqref{eq:mwi_implicit} reduces the errors associated with a discontinuous change in density \citep{Bartholomew2018}.}

The volume flux $F_f^{(n+1)}=\vartheta_f^{(n+1)} A_f$ is part of all governing equations and, thus, ensures the strong link between these equations, and prevents pressure-velocity decoupling as a consequence of the applied collocated variable arrangement.

\subsubsection{Calculation of the colour function $\widetilde{\psi}_f^{(n)}$}
\label{subsubsection:THINC_QQ}

The THINC/QQ interface-capturing scheme, a state-of-the-art version of the original THINC scheme of~\citet{Xiao2005,Xiao2011}, was first introduced by~\citet{Xie2017}, and recently revisited by~\citet{Chen2022b}. It relies on the use of a smoothed phase indicator $\tilde{H}$ to reconstruct the colour function $\psi$ at face $f$. This indicator is defined using a trigonometric function, and reconstructed within each cell $P$ as
\begin{equation}
    \tilde{H}_P(\mathbf{X}) = \frac{1}{2}\left[1+\tanh(\beta_P(\mathcal{P}_P(\mathbf{X})+d_P))\right],
    \label{eq:THINC_tanh_def}
\end{equation}
where $\beta_P$ is a local steepness parameter, computed in the present study from a user-defined scaling factor $\beta_f$, which is typically set to 6~\citep{Chen2022b}, and the local cell size $\Delta s_P = \sqrt[3]{V_P}$:
\begin{equation}
    \beta_P = \frac{\beta_f}{\Delta s_P}.
    \label{eq:beta_def}
\end{equation}
Thus, the smooth phase indicator $\tilde{H}$ relates to the sharp indicator $H$ as
\begin{equation}
    \lim_{\beta\to\infty} \tilde{H} = H,
    \label{eq:THINC_lim_beta}
\end{equation}
$\mathbf{X}$ are the local cell-wise coordinates, which can be defined for a Cartesian cell $P$ of dimensions ($\Delta x_P, \Delta y_P, \Delta z_P$) as
\begin{equation}
    \mathbf{X} = \begin{pmatrix}
           X \\
           Y \\
           Z 
         \end{pmatrix}
         = \begin{pmatrix}
           \dfrac{2 (x - x_P)}{\Delta x_P} \\
           \dfrac{2 (y - y_P)}{\Delta y_P}\\
           \dfrac{2 (z - z_P)}{\Delta z_P}
         \end{pmatrix},
    \label{eq:local_coords_def}
\end{equation}
with $\mathbf{x}_P = (x_P, y_P, z_P)$ the coordinates of the cell center. Hence, $-1\leq X, Y, Z \leq 1$ and $\mathbf{X}_0 = (0, 0, 0)$ corresponds to the center of the computational cell. The quadratic interface polynomial $\mathcal{P}$, which contains the local geometric information of the interface (normal vector, curvature), is defined as
\begin{equation}
    \mathcal{P}_P(\mathbf{X}) = C_{200,P} X^2 + C_{020,P} Y^2 + C_{002,P} Z^2
	+ C_{110,P} XY + C_{101,P} XZ + C_{011,P} YZ
	+ C_{100,P} X + C_{010,P} Y + C_{001,P} Z	,
    \label{eq:THINC_polynomial_def}
\end{equation}
and $d_P$ is the local surface constant, which sets the position of the interface within a cell $P$ containing the interface, computed by enforcing volume conservation for each cell as
\begin{equation}
	\frac{1}{V_P}\int_V \tilde{H}_P(\mathbf{X}) \, \mathrm{d}V = \psi_P.
    \label{eq:d_cst_def}
\end{equation}

The quadratic polynomial coefficients $C$ in Eq.~\eqref{eq:THINC_polynomial_def} embed the local geometric information of the cell-wise parabolic interface, as described in detail by \citet{Xie2017} and \citet{Chen2022b}. \textcolor{black}{They are computed in local coordinates for each cell $P$, from the colour-function gradient normalised in global coordinates, as
\begin{equation}\label{eq:THINC_pol_coeffs_def}
\begin{split}
    C_{200,P} &= \mathcal{K}_{\mathrm{XX},P} / 2 = \left(\mathcal{H}_{\mathrm{xx},P} \left(\Delta x_P / 2\right)^2\right)/2     \\
    C_{020,P} &= \mathcal{K}_{\mathrm{YY},P} / 2 = \left(\mathcal{H}_{\mathrm{yy},P} \left(\Delta y_P / 2\right)^2\right)/2         \\
    C_{002,P} &= \mathcal{K}_{\mathrm{ZZ},P} / 2 =  \left(\mathcal{H}_{\mathrm{zz},P} \left(\Delta z_P / 2\right)^2\right)/2        \\
    C_{110,P} &= \mathcal{K}_{\mathrm{XY},P} =  \mathcal{H}_{\mathrm{xy},P}/2 \left(\Delta y_P / 2\right)^2 +  \mathcal{H}_{\mathrm{yx},P}/2 \left(\Delta x_P / 2\right)^2       \\
    C_{011,P} &= \mathcal{K}_{\mathrm{YZ},P} =  \mathcal{H}_{\mathrm{yz},P}/2 \left(\Delta z_P / 2\right)^2 +  \mathcal{H}_{\mathrm{zy},P}/2 \left(\Delta y_P / 2\right)^2        \\
    C_{101,P} &= \mathcal{K}_{\mathrm{XZ},P} = \mathcal{H}_{\mathrm{xz},P}/2 \left(\Delta z_P / 2\right)^2 +  \mathcal{H}_{\mathrm{zx},P}/2 \left(\Delta x_P / 2\right)^2         \\
    C_{100,P} &= \mathcal{N}_{\mathrm{x},P} \left(\Delta x_P/2\right)          \\
    C_{010,P} &= \mathcal{N}_{\mathrm{y},P} \left(\Delta y_P/2\right)           \\
    C_{001,P} &= \mathcal{N}_{\mathrm{z},P} \left(\Delta z_P/2\right),
\end{split}
\end{equation}
where $\boldsymbol{\mathcal{N}} = (\mathcal{N}_{\mathrm{x}},\,\mathcal{N}_{\mathrm{y}}, \,\mathcal{N}_{\mathrm{z}})$ is the unit normal vector in global coordinates $(x,y,z)$,
\begin{equation}
    \boldsymbol{\mathcal{N}} = \frac{\boldsymbol{\nabla}\psi}{|\boldsymbol{\nabla}\psi|},
    \label{eq:normal_vec_global}
\end{equation}
$\boldsymbol{\mathcal{H}} = \boldsymbol{\nabla}\boldsymbol{\mathcal{N}}$ is the Hessian tensor in global coordinates $(x,y,z)$,
and $\boldsymbol{\mathcal{K}}$ is the curvature tensor in local coordinates $(X,Y,Z)$.
In order to increase the accuracy of the computation of $\boldsymbol{\mathcal{N}}$, $5$ steps of Laplace smoothing are performed on the colour function before (and only for) the present normalised gradient computation.}

The THINC-based procedures are divided into two main steps: (i) algebraic reconstruction of the smooth phase indicator $\tilde{H}_P(\mathbf{X})$ for each cell, i.e.~computation of $\mathcal{P}$ and $d$, and (ii) calculation of the flux to advect the colour function $\psi_P$ using $\tilde{H}_P(\mathbf{X})$. Contrary to the other THINC schemes, the THINC/QQ procedure uses both a quadratic surface representation defined by Eq.~\eqref{eq:THINC_polynomial_def}, as well as a Gauss-quadrature integration of $\tilde{H}_P(\mathbf{X})$ for the calculation of the surface constant $d$ as defined by Eq.~\eqref{eq:d_cst_def} and for the advection flux.

The local surface constant $d_P$ is computed by enforcing volume conservation as indicated by Eq.~\eqref{eq:d_cst_def}, which features the volume integral of the hyperbolic tangent profile $\tilde{H}_P(\mathbf{X})$. Since no analytical solution exists to perform the multidimensional integration, a fully multidimensional Gaussian quadrature is employed to approximate the integral, as proposed by~\citet{Xie2017}, which greatly simplifies the numerical procedure.
Let us define $\mathbf{X}_q$ and $\omega_q$ ($q=1,...,Q$) the quadrature-point local coordinates and associated weights, respectively. In this work, the quadrature points and weights of~\citet{Xie2017} for hexahedral grids ($Q=16$ points) are used. Eq.~\eqref{eq:d_cst_def} is then given as
\begin{equation}
    \sum_{q=1}^Q \omega_{q,P} \left(\tilde{H}_P\left(\mathbf{X}_{q,P}\right)\right) = \sum_{q=1}^Q \omega_{q,P} \left(\frac{1}{2}\left(1+\tanh\left(\beta_P\left(\mathcal{P}_P\left(\mathbf{X}_{q,P}\right)+d_P\right)\right)\right)\right) = \psi_P,
    \label{eq:d_cst_calc_1}
\end{equation}
where the weights are defined so that $\sum_{q=1}^Q \omega_{q,P} = 1$~\citep{Xie2017}. Using the identity
\begin{equation}
    \tanh\left(\beta_P\mathcal{P}_P\left(\mathbf{X}_{q,P}\right)+\beta_P d_P\right) = \frac{\tanh\left(\beta_P\mathcal{P}_P\left(\mathbf{X}_{q,P}\right)\right)+\tanh\left(\beta_P d_P\right)}{1+\tanh\left(\beta_P\mathcal{P}_P\left(\mathbf{X}_{q,P}\right)\right)\cdot\tanh\left(\beta_P d_P\right)},
    \label{eq:d_cst_trigo_identity}
\end{equation}
Eq.~\eqref{eq:d_cst_calc_1} can be rewritten as
\begin{equation}
     \sum_{q=1}^Q \omega_{q,P} \frac{\tanh\left(\beta_P\mathcal{P}_P\left(\mathbf{X}_{q,P}\right)\right)+\tanh\left(\beta_P d_P\right)}{1+\tanh\left(\beta_P\mathcal{P}_P\left(\mathbf{X}_{q,P}\right)\right)\cdot\tanh\left(\beta_P d_P\right)} = 2\left(\psi_P - \frac{1}{2}\right).
    \label{eq:d_cst_calc_2}
\end{equation}
Provided the quadratic function $\mathcal{P}(\mathbf{X})$ is already known using Eq.~\eqref{eq:THINC_polynomial_def}, $d_P$ is the only remaining unknown to solve. Following~\citet{Xie2017}, $A_{q,P} = \tanh\left(\beta_P\mathcal{P}_P\left(\mathbf{X}_{q,P}\right)\right)$, $D_P = \tanh\left(\beta_P d_P\right)$, and $Q_P = 2\left(\psi_P - 1/2\right)$, which allows to write Eq.~\eqref{eq:d_cst_calc_2} as
\begin{equation}
     \sum_{q=1}^Q \omega_{q,P} \frac{A_{q,P} + D_P}{1+ A_{q,P} D_P} - Q_P = 0,
    \label{eq:d_cst_calc_3}
\end{equation}
which is a rational equation for the modified surface constant $D_P\in[-1,1]$. A Newton-Raphson algorithm is employed to find $D_P$, as described by~\citet{Kumar2021}.

Once the reconstruction of $\tilde{H}_P(\mathbf{X})$ has been performed for all interface cells (i.e.~cells with $10^{-8}<\psi_P<1-10^{-8}$), the associated flux can be computed to advect the colour function using Eq.~\eqref{eq:colour_advection}. As mentioned above, this flux involves the quantity $\tilde{\psi}_f^{(n)}$, which is computed in the THINC/QQ framework using Gauss-quadrature integration over the cell face $f$ as~\citep{Xie2017}
\begin{equation}
    \tilde{\psi}_f^{(n)} = \frac{1}{A_f}\int_{A_f}\tilde{H}_{P,\mathrm{upw}}^{(n)}\mathrm{d}A = \sum_{q=1}^{\mathcal{G}} \omega_q \tilde{H}_{P,\mathrm{upw}}^{(n)}(\mathbf{X}_q),
    \label{eq:THINC_QQ_flux}
\end{equation}
where $\tilde{H}_{P,\mathrm{upw}}^{(n)}$ is the reconstructed smooth phase indicator at nonlinear iteration $n$ in the upwind cell associated with face $f$, defined as
\begin{equation}
\tilde{H}_{P,\mathrm{upw}}^{(n)} =
	\begin{cases}
	 \tilde{H}_{P}^{(n)} \,\, \text{if} \,\,  \vartheta_f^{(n)} > 0 \\
	 \tilde{H}_{Q}^{(n)} \,\, \text{if} \,\,  \vartheta_f^{(n)} < 0.
	\end{cases}
    \label{eq:THINC_QQ_upw_tanh}
\end{equation}
For the Gauss-quadrature integration, $\mathcal{G}=9$ Gauss-Legendre points are used in total on the cell-face surface. With the colour function at cell face $f$ defined by Eq.~\eqref{eq:THINC_QQ_flux}, the colour function flux follows as $\mathcal{F}_f^{(\psi)} = \widetilde{\psi}_f^{(n)} F_f^{(n+1)}$.

\subsubsection{Calculation of density and momentum fluxes}
\label{subsubsection:flux_density_momentum}

Following \citet{Arrufat2021}, the density flux is formulated consistently with the colour-function flux as
\begin{equation}
    \mathcal{F}_f^{(\rho)} = \rho_\mathrm{A} \mathcal{F}_f^{(\psi)} + \rho_\mathrm{B} \mathcal{F}_f^{(1-\psi)} = \rho_\mathrm{A} \widetilde{\psi}_f^{(n)}F_f^{(n+1)} + \rho_\mathrm{B} \left(1-\widetilde{\psi}_f^{(n)}\right)F_f^{(n+1)}.
    \label{eq:density_flux_colour_dependency}
\end{equation}
Rearranging this equation gives
\begin{equation}
    \mathcal{F}_f^{(\rho)} = \left[\rho_\mathrm{B} + \widetilde{\psi}_f^{(n)}\left(\rho_\mathrm{A}-\rho_\mathrm{B}\right)\right]F_f^{(n+1)} = \widetilde{\rho}_f^{(n)} F_f^{(n+1)},
    \label{eq:density_flux_colour_final}
\end{equation}
where the face density at the nonlinear iteration $n$ is
\begin{equation}
    \widetilde{\rho}_f^{(n)} = \rho_\mathrm{B} + \widetilde{\psi}_f^{(n)}\Delta\rho
    \label{eq:face_density}
\end{equation}
and $\Delta \rho = \rho_\text{A}-\rho_\text{B}$.
The momentum flux follows, in a similar manner, based on the density flux as
\begin{equation}
    \mathcal{F}_f^{(\rho\mathbf{u})} = \widetilde{\rho}_f^{(n)} \widetilde{\mathbf{u}}_f^{(n)}  F_f^{(n+1)} = \mathcal{F}_f^{(\rho)}\widetilde{\mathbf{u}}_f^{(n)}.
    \label{eq:momentum_flux_1}
\end{equation}

The only quantity that remains to be calculated is the face velocity $\widetilde{\mathbf{u}}_f^{(n)}$. To this end, we employ a Favre averaging together with total variation diminishing (TVD) differencing, following the work of~\citet{Kuhn2023}, such that the velocity at face $f$ is defined as
\begin{equation}
    \widetilde{\mathbf{u}}_f^{(n)} = \frac{(\rho\mathbf{u})_{f,\mathrm{TVD}}^{(n)}}{\rho_{f,\mathrm{TVD}}^{(n)}},
    \label{eq:advected_vel}
\end{equation}
where the TVD quantities $\varphi$ are calculated using the Minmod flux limiter $\xi_{f,\mathrm{Minmod}}^{(\varphi)}$~\citep{Roe1986,  Denner2015a} as
\begin{align}
    \rho_{f,\mathrm{TVD}}^{(n)} & = \rho_U^{(n)} + \xi_{f,\mathrm{Minmod}}^{(\rho)}\left[\rho_D^{(n)} - \rho_U^{(n)}\right] \\
    (\rho\mathbf{u})_{f,\mathrm{TVD}}^{(n)} & = (\rho\mathbf{u})_U^{(n)} + \xi_{f,\mathrm{Minmod}}^{(\rho\mathbf{u})}\left[(\rho\mathbf{u})_D^{(n)} - (\rho\mathbf{u})_U^{(n)}\right].
\end{align}
The subscripts $U$ and $D$ denote the upwind and downwind cells, respectively.

\subsection{Discretisation of transient terms}
\label{subsection:transient_terms_discretisation}

A second-order backward Euler time-integration scheme for variable time steps is employed to discretise the transient terms, given for the general fluid variable $\Omega$ as
\citep{Moukalled2016}
\begin{equation}
    \int_{V_P} \frac{\partial \Omega}{\partial t} \, \text{d}V \approx \left[ \left(\frac{1}{\Delta t}+\frac{1}{\Delta \tau}\right) \Omega_P^{(n+1)} - \left(\frac{1}{\Delta t}+\frac{1}{\Delta t^\mathrm{o}}\right) \Omega_P^{(t-\Delta t)}+ \frac{\Delta t}{\Delta t^\mathrm{o} \Delta \tau} \, \Omega_P^{(t-\Delta \tau)} \right] V_P, \label{eq:bdf2}
\end{equation}
where $\Delta t$ is the current time step, $\Delta t^\mathrm{o}$ denotes the previous time step and $\Delta \tau = \Delta t+\Delta t^\mathrm{o}$.
The transient term of the advection equation of the colour function, Eq.~\eqref{eq:colour_advection}, is readily discretised by Eq.~\eqref{eq:bdf2} with $\Omega = \psi$. Using the cell-based density computed from the cell-based colour function as $\rho_P = \rho_\mathrm{B} + \Delta\rho \,\psi_P$, where $\Delta \rho=\rho_\mathrm{A}-\rho_\mathrm{B}$, the implicit cell-based density is obtained by performing a Newton linearisation as
\begin{equation}
    \rho_P^{{(n+1)}} =  \rho_P^{{{(n)}}} + \left(\psi_P^{{{(n+1)}}} - \psi_P^{{{(n)}}}\right)\frac{\partial \rho}{\partial \psi} =\rho_P^{(n)} + \left(\psi_P^{{{(n+1)}}} - \psi_P^{{{(n)}}}\right) \Delta\rho.
    \label{eq:newton_density_wrt_colour}
\end{equation}
The transient term of the continuity equation is then discretised using Eq.~\eqref{eq:bdf2} with $\Omega_P^{(n+1)}=\rho_P^{{(n+1)}}$ given by Eq.~\eqref{eq:newton_density_wrt_colour}, $\Omega_P^{(t-\Delta t)} = \rho_P^{(t-\Delta t)}$, and $\Omega_P^{(t-\Delta \tau)} = \rho_P^{(t-\Delta \tau)}$. The implicit momentum is defined as a product of density and velocity, linearised with a Newton linearisation, 
\begin{equation}
    (\rho\mathbf{u})_P^{(n+1)} = \rho_P^{(n+1)}\mathbf{u}_P^{(n)} + \rho_P^{(n)}\mathbf{u}_P^{(n+1)} - \rho_P^{(n)}\mathbf{u}_P^{(n)}.
    \label{eq:momentum_transient_2}
\end{equation}
Rearranging the terms of this equation and inserting the density definition used for the discretisation of the transient term of the continuity equation yields
\begin{equation}
    (\rho\mathbf{u})_P^{(n+1)} = \rho_P^{(n)}\mathbf{u}_P^{(n+1)} + \left(\psi_P^{(n+1)}-\psi_P^{(n)}\right) \Delta \rho \, \mathbf{u}_P^{(n)}.
    \label{eq:momentum_transient_3}
\end{equation}
The transient term of the momentum equations is then discretised using Eq.~\eqref{eq:bdf2} with $\Omega_P^{(n+1)}=(\rho\mathbf{u})_P^{(n+1)}$ given by Eq.~\eqref{eq:momentum_transient_3}, $\Omega_P^{(t-\Delta t)} =  (\rho\mathbf{u})_P^{(t-\Delta t)}$, and $\Omega_P^{(t-\Delta \tau)} = (\rho\mathbf{u})_P^{(t-\Delta \tau)}$.

\subsection{Discretised equation system and solution procedure}
\label{subsection:full_syst_discretisation}

Applying the discretisation defined in Sections \ref{subsection:discretisation_surf_tens}-\ref{subsection:transient_terms_discretisation}, the discretised continuity equation, Eq.~\eqref{eq:NS_mass_cons}, for mesh cell $P$ is given as
\begin{equation}
{\left[ \left(\frac{1}{\Delta t}+\frac{1}{\Delta \tau}\right) \left[\rho_P^{(n)} + \left(\psi_P^{{{(n+1)}}} - \psi_P^{{{(n)}}}\right) \Delta\rho\right] - \left(\frac{1}{\Delta t}+\frac{1}{\Delta t^\mathrm{o}}\right) \rho_P^{(t-\Delta t)}+ \frac{\Delta t}{\Delta t^\mathrm{o} \Delta \tau} \, \rho_P^{(t-\Delta \tau)} \right] V_P} + {\sum_f \mathcal{F}_f^{(\rho)}} = 0,
\label{eq:continuity_disc}
\end{equation}
the discretised momentum equations, Eq.~\eqref{eq:NS_momentum_cons}, follows as
\begin{equation}
    \begin{split}
        & {\left[ \left(\frac{1}{\Delta t}+\frac{1}{\Delta \tau}\right) \left[\rho_P^{(n)}\mathbf{u}_P^{(n+1)} + \left(\psi_P^{(n+1)}-\psi_P^{(n)}\right) \Delta \rho \, \mathbf{u}_P^{(n)}\right] - \left(\frac{1}{\Delta t}+\frac{1}{\Delta t^\mathrm{o}}\right) (\rho\mathbf{u})_P^{(t-\Delta t)}+ \frac{\Delta t}{\Delta t^\mathrm{o} \Delta \tau} \, (\rho\mathbf{u})_P^{(t-\Delta \tau)} \right] V_P} \\ &+ {\sum_f \mathcal{F}^{(\rho\mathbf{u})}_f} = {- \sum_f \overline{p}_f^{\textcolor{black}{(n+1)}} \mathbf{n}_{f} A_f} + { \sum_f \mu_f \left( \frac{\mathbf{u}_{Q}^{\textcolor{black}{(n+1)}} - \mathbf{u}_{P}^{\textcolor{black}{(n+1)}}}{\Delta x} \,  + \overline{\boldsymbol{\nabla}\mathbf{u}}_f^{\textcolor{black}{(n+1)}}\cdot\mathbf{n}_{f}\right)  A_f} + {{\mathbf{S}}_{\sigma,P}^{\textcolor{black}{(n+1)}} V_P},
    \end{split}
    \label{eq:momentum_disc}
\end{equation}
and the discretised advection equation for the colour function, Eq.~~\eqref{eq:colour_advection}, is given as
\begin{equation}
    {\left[ \left(\frac{1}{\Delta t}+\frac{1}{\Delta \tau}\right) \psi_P^{(n+1)} - \left(\frac{1}{\Delta t}+\frac{1}{\Delta t^\mathrm{o}}\right) \psi_P^{(t-\Delta t)}+ \frac{\Delta t}{\Delta t^\mathrm{o} \Delta \tau} \, \psi_P^{(t-\Delta \tau)} \right] V_P} + {\sum_f \mathcal{F}^{(\psi)}_f} - {\psi_P^{\textcolor{black}{{(n)}}}\sum_f F_f^{\textcolor{black}{{(n+1)}}}} = 0,
    \label{eq:vof_disc}
\end{equation}
where $\overline{\square}_f$ denotes a linear interpolation of the cell-centred values to face $f$.
This discretised set of governing equations is solved simultaneously for the pressure $p$, the velocity vector $\mathbf{u} = (u~v~w)^\text{T}$ and the colour function $\psi$ in a single linear equation system, $\boldsymbol{\mathcal{A}} \cdot \boldsymbol{\phi} = \mathbf{b}$, given for a three-dimensional mesh with $N$ cells as 
\begin{equation}
    \begin{pmatrix}
        \boldsymbol{\mathcal{A}}^{p}_\text{cont.} 
        & \boldsymbol{\mathcal{A}}^u_\text{cont.} 
        & \boldsymbol{\mathcal{A}}^v_\text{cont.} 
        & \boldsymbol{\mathcal{A}}^w_\text{cont.}
        & \boldsymbol{\mathcal{A}}^\psi_\text{cont.} \\
        \boldsymbol{\mathcal{A}}^{p}_\text{$x$-mom.} 
        & \boldsymbol{\mathcal{A}}^u_\text{$x$-mom.} 
        & \boldsymbol{\mathcal{A}}^v_\text{$x$-mom.} 
        & \boldsymbol{\mathcal{A}}^w_\text{$x$-mom.}
        & \boldsymbol{\mathcal{A}}^\psi_\text{$x$-mom.} \\
        \boldsymbol{\mathcal{A}}^{p}_\text{$y$-mom.} 
        & \boldsymbol{\mathcal{A}}^u_\text{$y$-mom.} 
        & \boldsymbol{\mathcal{A}}^v_\text{$y$-mom.} 
        & \boldsymbol{\mathcal{A}}^w_\text{$y$-mom.}
        & \boldsymbol{\mathcal{A}}^\psi_\text{$y$-mom.} \\
        \boldsymbol{\mathcal{A}}^{p}_\text{$z$-mom.} 
        & \boldsymbol{\mathcal{A}}^u_\text{$z$-mom.} 
        & \boldsymbol{\mathcal{A}}^v_\text{$z$-mom.} 
        & \boldsymbol{\mathcal{A}}^w_\text{$z$-mom.}
        & \boldsymbol{\mathcal{A}}^\psi_\text{$z$-mom.} \\
        \boldsymbol{\mathcal{A}}^{p}_\text{\sc vof} 
        & \boldsymbol{\mathcal{A}}^u_\text{\sc vof} 
        & \boldsymbol{\mathcal{A}}^v_\text{\sc vof} 
        & \boldsymbol{\mathcal{A}}^w_\text{\sc vof}
        & \boldsymbol{\mathcal{A}}^\psi_\text{\sc vof} \\
    \end{pmatrix}
    \cdot
    \begin{pmatrix}
        \boldsymbol{\phi}^p \\ 
        \boldsymbol{\phi}^u \\ 
        \boldsymbol{\phi}^v \\ 
        \boldsymbol{\phi}^w \\
        \boldsymbol{\phi}^\psi
    \end{pmatrix}
    =
    \begin{pmatrix}
        \mathbf{b}_\text{cont.} \\ 
        \mathbf{b}_\text{$x$-mom.} \\ 
        \mathbf{b}_\text{$y$-mom.} \\ 
        \mathbf{b}_\text{$z$-mom.} \\
        \mathbf{b}_\text{\sc vof}
    \end{pmatrix}.
    \label{eq:eqsysfull}
\end{equation}
The coefficient submatrices of size $N \times N$ for each governing equation ``eq.'' with regard to each solution variable $\chi \in \{p,u,v,w,\psi\}$ are denoted as $\boldsymbol{\mathcal{A}}^\chi_\text{eq.}$, where ``eq.~= cont.'' refers to the continuity equation, Eq.~\eqref{eq:continuity_disc}, $\mathrm{eq.}= \{x\text{-mom.}, y\text{-mom.}, z\text{-mom.}\}$  refers to the three momentum equations with respect to each spatial dimension, Eq.~\eqref{eq:momentum_disc}, and ``eq.~= {\sc vof}'' refers to the advection equation of the VOF colour function, Eq.~\eqref{eq:vof_disc}.  $\boldsymbol{\phi}^\chi$ are the solution subvectors of length $N$ for each solution variable $\chi$ and all contributions from previous time-levels and contributions that are deferred are contained in the right-hand side subvectors $\mathbf{b}_\text{eq.}$, each of length $N$. 

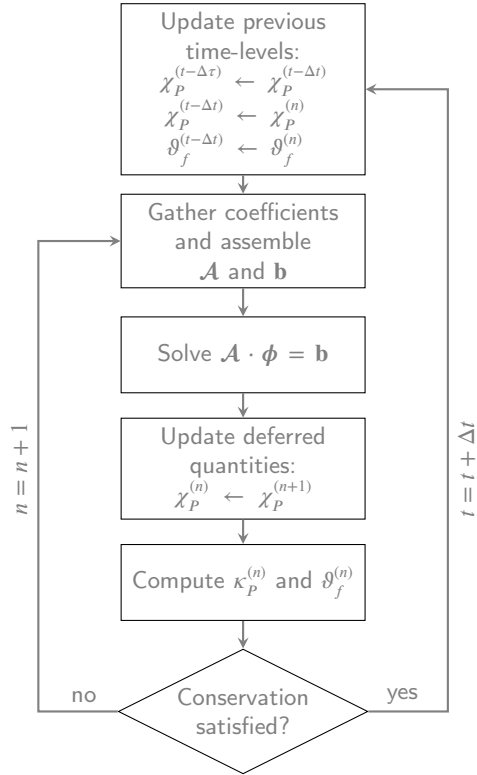
\begin{figure}[t]
\begin{center}
\begin{small}
    \begin{tikzpicture}
    \node (pro0) [process] {Update \mbox{previous} time-levels: $\chi_P^{(t-\Delta \tau)}\leftarrow \chi_P^{(t-\Delta t)}$ $\chi_P^{(t-\Delta t)}\leftarrow \chi_P^{(n)\phantom{...)}}$ $\vartheta_f^{(t-\Delta t)} \leftarrow \vartheta_f^{(n)\phantom{....}}$};
    \node (pro1a) [process, below of=pro0, yshift=-1cm] {\mbox{Gather coefficients} \mbox{and assemble}  \mbox{$\boldsymbol{\mathcal{A}}$ and $\mathbf{b}$}}; 
    \node (pro1b) [process, below of=pro1a, yshift=-0.5cm] {Solve $\boldsymbol{\mathcal{A}} \cdot \boldsymbol{\phi} = \mathbf{b}$}; 
    \node (pro2) [process, below of=pro1b, yshift=-0.5cm] {Update \mbox{deferred} quantities: $\chi_P^{(n)}
    \leftarrow \chi_P^{(n+1)}$};
    % \node (pro3) [process, below of=pro2, yshift=-0.5cm] {Compute $\kappa^{(n)}$};
    % \node (pro4) [process, below of=pro3, yshift=-0.4cm] {Compute $\vartheta_f^{(n)}$};
    \node (pro3) [process, below of=pro2, yshift=-0.5cm] {Compute $\kappa_P^{(n)}$ and $\vartheta_f^{(n)}$};
    \node (dec1) [decision, below of=pro3, yshift=-0.7cm]
    {Conservation satisfied?}; 
    \draw [arrow] (pro0) -- (pro1a);
    \draw [arrow] (pro1a) -- (pro1b);
    \draw [arrow] (pro1b) -- (pro2);
    \draw [arrow] (pro2) -- (pro3);
    % \draw [arrow] (pro3) -- (pro4);
    \draw [arrow] (pro3) -- (dec1);
    \draw [arrow] (dec1) --+(-2.7cm,0) |- (pro1a);
    \draw [arrow] (dec1) --+(+2.7cm,0) |- (pro0);
    \node at (-2.1,-8) {no};
    \node at (2.1,-8) {yes};
    \node [rotate=90] at (-2.95,-5) {$n=n+1$};
    \node [rotate=90] at (2.95,-5) {$t=t+\Delta t$};
    \end{tikzpicture} 
\end{small}
\caption{Flow chart of the solution procedure of the discretised and linearised fully-coupled system of governing equations, $\boldsymbol{\mathcal{A}} \cdot \boldsymbol{\phi} = \mathbf{b}$. Superscript $(n+1)$ denotes implicitly solved variables,  superscript $(n)$ denotes deferred variables, and superscripts $(t-\Delta t)$ and $(t-\Delta \tau)$ denote quantities of previous time-levels.}
\label{fig:flowchart}
\end{center}
\end{figure}

The linear system of discretised governing equations, Eq.~\eqref{eq:eqsysfull}, is solved iteratively using the Block-Jacobi pre-conditioner and the BiCGSTAB solver of the software library PETSc~\citep{petsc-user-ref,  petsc-web-page}. To account for the nonlinearity of the governing equations, an inexact Newton method~\citep{Dembo1982} is applied, whereby the deferred terms are updated iteratively until the nonlinear system of equations has converged to a predefined conservation criteria, as illustrated in Fig.~\ref{fig:flowchart}. The nonlinear iterative process is considered to be converged when the maximum $L_2$ error norm of the quantities $\phi\in\{\rho, \rho u, \rho v, \rho w, \psi\}$ satisfies
\begin{equation}
    \max_\phi\left(L_2(\phi)\right) < \epsilon_\text{nonlinear},
    \label{eq:nonlinear_convergence}
\end{equation}
where $\epsilon_\text{nonlinear}$ is typically set to $10^{-6}$, and $L_2(\phi)$ is calculated as
\begin{equation}
    L_2(\phi) = \sqrt{\frac{1}{N_\text{cells}}\frac{\sum_P\left(\phi_P^{(n+1)}-\phi_P^{(n)}\right)^2}{\max_{P}\left(\left(\phi_P^{(n+1)}\right)^2\right)}}.
    \label{eq:nonlinear_convergence_L2}
\end{equation}

As previously reported by \citet{Denner2022b}, since the implicit volume flux $F_f^{(n+1)}$ is based on implicit contributions of all solution variables $\chi \in \{p,u,v,w,\psi\}$ and contained in the advection terms of all governing equations, it introduces a tight implicit coupling between the governing equations, which was found to be essential for breaching the capillary time-step constraint. In the proposed algorithm, this implicit coupling between the governing equations is further strengthened by treating the density of the current time-level implicit with respect to the colour function in the transient terms of the continuity and momentum equations. Furthermore, to facilitate the robust solution of interfacial flows with large density ratios, the continuity and momentum equations are discretised in conservative form, contrary to the prevailing standard for incompressible flows, and a Favre averaging is applied in the momentum advection term. 

\section{Upper capillary stability limit}
\label{section:upper_limit}

Although the capillary time-step constraint has been shown to be breachable using interface-capturing methods,~\citet{Denner2022b} observed that, even beyond the capillary time-step constraint, the time step that yields a stable solution remains limited. Following the work of \citet{Galusinski2008}, the maximum possible time step using the class of fully-coupled algorithms considered in this study has the form \citep{Denner2022b}
\begin{equation}
    \Delta t_* = \frac{a_2\tau_\text{vc}+\sqrt{(a_2\tau_\text{vc})^2+4a_1\tau_\sigma^2}}{2},
    \label{eq:upper_dt_lim}
\end{equation}
where $\tau_\text{vc}$ and $\tau_\sigma$ are the visco-capillary and capillary time scales, respectively, and $a_{1,2}$ are two case-dependent constants. The capillary time-step constraint $\Delta t_\sigma$, Eq.~(\ref{eq:dt_sigma_theory_Denner}), is recovered for $a_1=1/(16\pi)$ and $a_2=0$.

By defining $\hat{\rho}=\rho_\mathrm{A}+\rho_\mathrm{B}$ and $\hat{\mu}=\mu_\mathrm{A}+\mu_\mathrm{B}$, and with the wavelength of the shortest unambiguously resolved capillary waves given as $\lambda_\sigma=2\Delta x$ \citep{Brackbill1992,Denner2015}, the reference time scales are  
\begin{align}
     \tau_\text{vc} & = \frac{\hat{\mu}\lambda_\sigma}{\sigma} \\
     \tau_\sigma    & = \sqrt{\frac{\hat{\rho}\lambda_\sigma^3}{\sigma}} ,
\end{align}
which yield the mesh Ohnesorge number \citep{Denner2022b}
\begin{equation}
    \mathrm{Oh}_{\Delta x} = \frac{\tau_\text{vc}}{\tau_\sigma} = \frac{\hat{\mu}}{\sqrt{\hat{\rho}\sigma\lambda_\sigma}}.
    \label{eq:upper_dt_lim_Oh_dx}
\end{equation}
The upper capillary stability limit can, therefore, be separated into two regimes: (i) the inviscid regime for $\mathrm{Oh}_{\Delta x}\ll 1$ where $\Delta t_* \propto \tau_\sigma$, and (ii) the viscous regime for $\mathrm{Oh}_{\Delta x}\gg 1$ where $\Delta t_* \propto \tau_\text{vc}$. In order to determine the case-dependent constants $a_1$ and $a_2$, the results of \citet{Denner2022b} suggest that only two results obtained for $\mathrm{Oh}_{\Delta x}\ll 1$ and $\mathrm{Oh}_{\Delta x}\gg 1$, respectively, are sufficient.

\section{Test cases and validation}
\label{section:validation}

The proposed fully-coupled algorithm is validated and scrutinized using four representative test cases in which surface tension plays the dominant role and for which the time step is usually limited by the capillary time-step constraint: the Laplace equilibrium of a stationary droplet (Section \ref{subsection:laplace_equilibrium}), a standing capillary wave (Section \ref{subsection:capillary_wave}), an oscillating droplet (Section \ref{subsection:oscillating_droplet}), and the Rayleigh-Plateau instability (Section \ref{subsection:rayleigh_plateau}). The analysis of these results focuses in particular on the force balancing of the discretised governing equations, the conservation of energy, and the maximum stable time step that can be applied in these simulations.

\subsection{Laplace equilibrium of a stationary droplet}
\label{subsection:laplace_equilibrium}

The common first step to validate a surface tension framework is to demonstrate its \textit{balanced-force} or \textit{well-balanced} property~\citep{Popinet2009,  Abadie2015,  Popinet2018,  Abu-Al-Saud2018} by simulating a static droplet (or bubble) and show that the exact Laplace balance $\Delta p = \sigma\kappa$ can be achieved, provided the interface shape has reached a numerical equilibrium. Given that the spherical interface should satisfy the Laplace balance and be in mechanical equilibrium, the two-phase flow should be quiescent. This is the case when the \textit{spurious currents}, which are the only source of a non-zero velocity $\mathbf{u}$, introduced by the initial numerical disequilibrium, are dissipated by viscosity. Hence, the relevant time scale to consider is the viscous dissipation time scale $\tau_\mu$, given for a droplet of diameter $D$ as $\tau_\mu = {\rho_\mathrm{A} D^2}/{\mu_\mathrm{A}}$. The relevant dimensionless parameter to parameterize this case is the Laplace number $\mathrm{La} = {\rho_\mathrm{A} \sigma D}/{\mu_\mathrm{A}^2}$. It is well-known that the exact Laplace equilibrium can be retrieved in segregated-VOF/explicit-CSF frameworks~\citep{Popinet2009}, and more recently with the coupled-VOF/implicit-CSF framework of ~\citet{Denner2022b} for time steps larger than the capillary time-step constraint for a uniform density ratio.
In this work, we consider a large density ratio of $\rho_\mathrm{A}/\rho_\mathrm{B}=1000$ for both a two-dimensional (2D) and three-dimensional (3D) droplet. The dimensionless parameters of the problem considered are summarised in Table~\ref{table_spurious_currents}.

\begin{table}[width=.9\linewidth,cols=5]
\caption{Dimensionless physical and numerical parameters for the Laplace equilibrium case.}\label{table_spurious_currents}
\begin{tabular*}{\tblwidth}{@{} LLLLL@{} }
\toprule
$\rho_\mathrm{A}/\rho_\mathrm{B}$ & $\mu_\mathrm{A}/\mu_\mathrm{B}$ & $\mathrm{La}$ & $D/\Delta x$ & $\Delta t/\Delta t_{\sigma}$\\
\midrule
1000 & 1 & 120 & 25.6 & 0.5, 2, 8, 16\\
\bottomrule
\end{tabular*}
\end{table}

\begin{figure}
\begin{minipage}[c]{.48\linewidth}
    \vspace{6pt}
    \centering
    \includegraphics[scale=0.55]{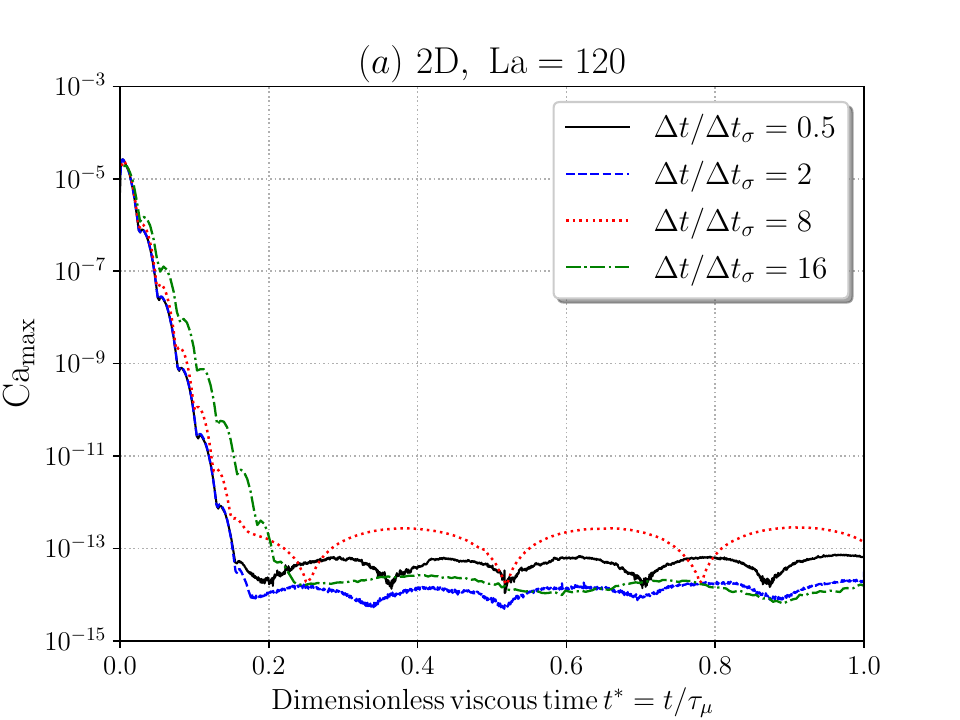}
       \centering
    \end{minipage}
    \hfill%
    \begin{minipage}[c]{.48\linewidth}
    \vspace{6pt}
      \centering
    \includegraphics[scale=0.55]{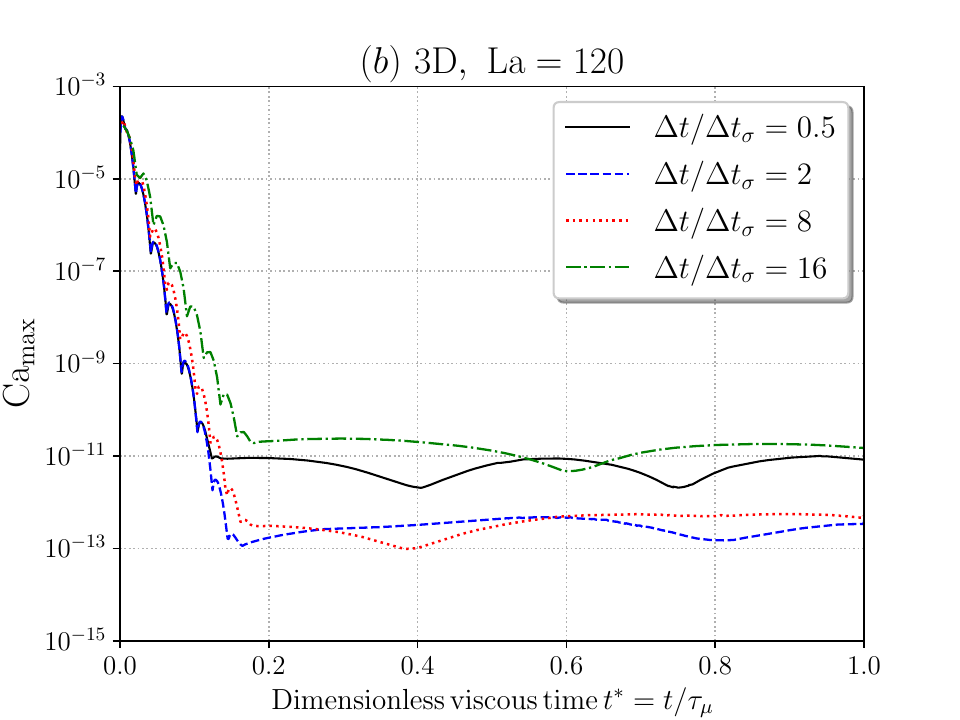}
       \centering
    \end{minipage}
    \caption{Maximum capillary number over time for the Laplace equilibrium case with density ratio $\rho_\text{A}/\rho_\text{B}=1000$ in (a) 2D and (b) 3D.}
    \label{fig:La_eq_Ca_over_time}
\end{figure}

The evolution of the maximum capillary number $\mathrm{Ca}_\mathrm{max} = {\mu_\mathrm{A} |\mathbf{u}|_\infty}/{\sigma}$ of the spurious currents as a function of the dimenonless time $t^* = t/\tau_\mu$ is shown in Fig.~\ref{fig:La_eq_Ca_over_time} (a) for the 2D case and in Fig.~\ref{fig:La_eq_Ca_over_time} (b) for the 3D case. In both cases, the initial spurious currents are dissipated exponentially by viscosity, until a numerical equilibrium is reached at $t^* \approx 0.2$. For $t^* > 0.2$, the magnitude of the spurious currents remains negligible and is defined by the tolerance to which the nonlinear system of governing equations is solved (or machine precision). 

\begin{figure}
\centering
		\includegraphics[scale=.175]{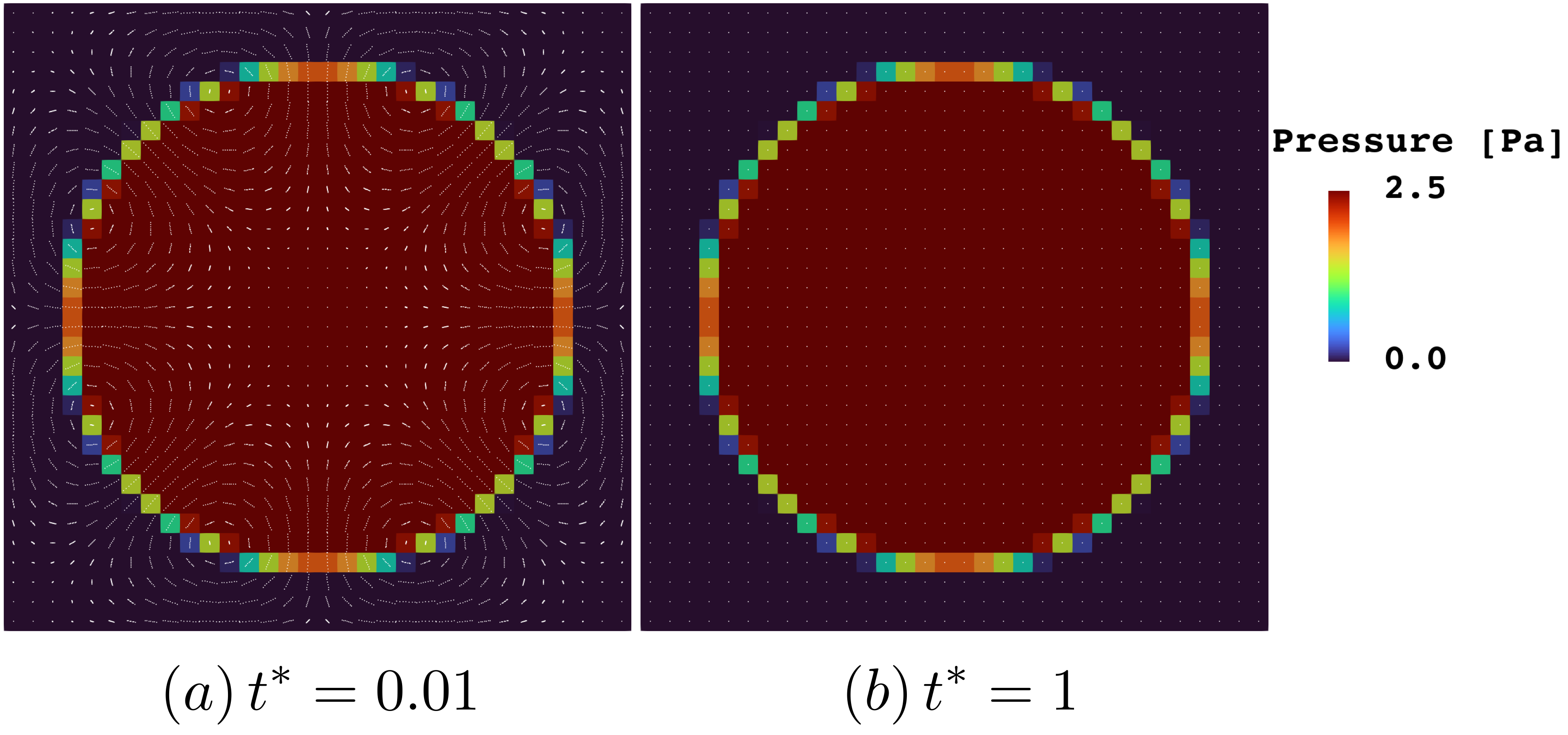}
		\caption{\textcolor{black}{Pressure field and spurious currents at (a) $t^*=t/\tau_\mu=0.01$ and (b) $t^*=t/\tau_\mu=1$ in 2D for $\Delta t/\Delta t_{\sigma}=16$.}}
		\label{fig:La_eq_contours}
\end{figure}

\textcolor{black}{The pressure field and spurious currents are illustrated in Fig.~\ref{fig:La_eq_contours}, at two time instants representing the beginning and the end of the 2D simulation with the largest applied time step, $\Delta t/\Delta t_\sigma=16$. The droplet has a radius of $R=0.4\mathrm{m}$ and the surface tension coefficient is $\sigma=1\mathrm{N}.\mathrm{m}^{-1}$, such that the pressure jump follows as $\Delta p=\sigma/R=2.5\mathrm{Pa}$, which is correctly predicted by the simulation, see Fig.~\ref{fig:La_eq_contours}.}

\begin{figure}
\centering
		\includegraphics[scale=.65]{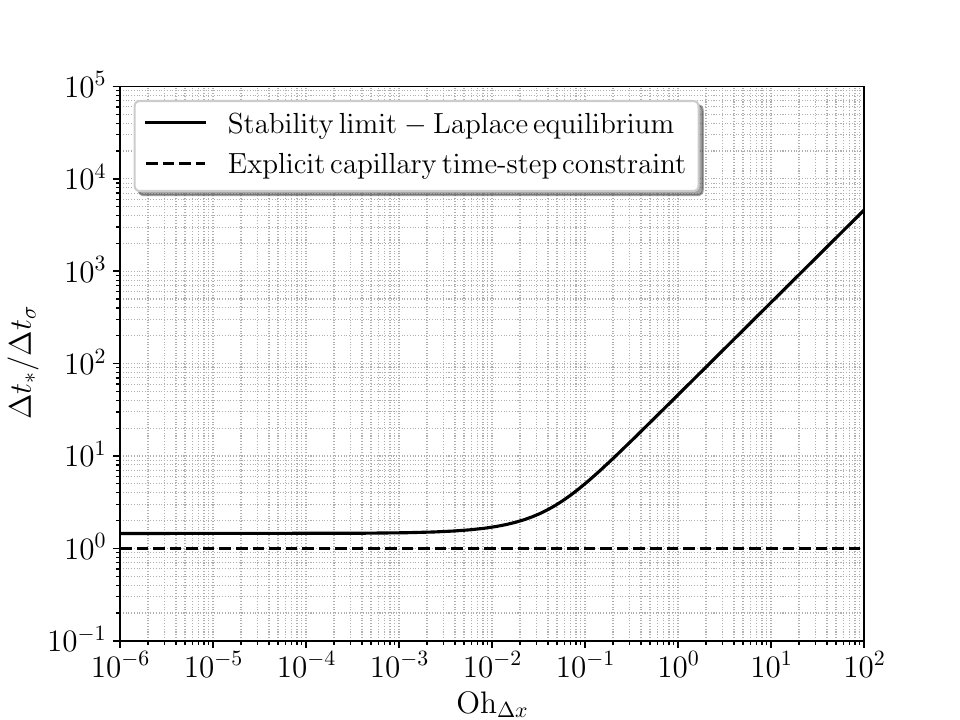}
		\caption{Approximated upper capillary stability limit as a function of the mesh Ohnesorge number, $\text{Oh}_{\Delta x}$, for the Laplace equilibrium case with density ratio $\rho_\text{A}/\rho_\text{B}=1000$.}
		\label{fig:upper_lim_La_eq}
\end{figure}

The upper capillary stability limit, discussed in Section~\ref{section:upper_limit}, for this case is displayed in Fig.~\ref{fig:upper_lim_La_eq}. In the limit of $\mathrm{Oh}_{\Delta x}\ll 1$, a time step smaller than $1.5 \Delta t_\sigma$ has to be applied to obtain a stable solution, which is significantly more restrictive than for the same case with unit density ratio for which a maximum time step of $15\Delta t_\sigma$ could be used in this regime \citep{Denner2022b}. In the large-$\mathrm{Oh}_{\Delta x}$ regime, the upper capillary stability limit $\Delta t_*$ is also one order of magnitude smaller than in the case with unit density ratio considered by~\citet{Denner2022b}. Nonetheless, the proposed fully-couppled algorithm is capable to breach the capillary time-step constraint even for interfacial flows with realistic gas-liquid density ratios and still retain the force balance between pressure and surface tension.

\subsection{Capillary wave}
\label{subsection:capillary_wave}

We consider the oscillation and decay of a capillary wave. Two immiscible viscous fluids with large density and viscosity ratios are initially at rest and separated by an interface with surface tension, perturbed by a small-amplitude sinusoidal capillary wave, in a two-dimensional $[0;\lambda]\times[0;3\lambda]$ domain \citep{Popinet1999, Denner2017a}. This domain is periodic in the $x$-direction and has slip walls at its top and bottom boundaries. The initial wave amplitude is $A_0=\lambda/100$, with $\lambda=2\pi$ the length of the capillary wave, which gradually decays due to viscous dissipation. The calculations are performed on Cartesian meshes with a mesh resolution of $\lambda/\Delta x=\{25,50,100,200\}$, using different time steps $\Delta t/\Delta t_\sigma=\{0.5, 2, 8\}$, and run until $\omega_0 t = 25$, which corresponds to approximately $4$ oscillations and where $\omega_0 = \sqrt{\sigma k^3/(\rho_\mathrm{A}+\rho_\mathrm{B})}$ is the undamped angular frequency of the capillary wave. The considered density and viscosity ratios are $\rho_\text{A}/\rho_\text{B}=1000$ and $\mu_\text{A}/\mu_\text{B}=1000$, such that the kinematic viscosity $\nu=\mu/\rho$ is $\nu_\text{A} = \nu_\text{B}$. The physical and numerical parameters of this case are summarised in Table~\ref{table_2D_damped_wave_1}. The results obtained with the proposed fully-coupled algorithm are compared to the analytical solution of the temporal evolution of the wave amplitude proposed by \citet{Prosperetti1981}. This analytical solution is valid for capillary waves with small amplitude ($A\ll \lambda$) and for interacting fluids with equal kinematic viscosity $\nu$.

\begin{table}[width=.9\linewidth,cols=6]
\caption{Dimensionless physical and numerical parameters for the case of the viscous damping of a capillary wave.}
\label{table_2D_damped_wave_1}
\begin{tabular*}{\tblwidth}{@{} LLLLL@{} }
\toprule
$\rho_\mathrm{A}/\rho_\mathrm{B}$ & $\mu_\mathrm{A}/\mu_\mathrm{B}$ & $\mathrm{La}=\rho\lambda\sigma/\mu^2$ & $\lambda/\Delta x$ & $\Delta t/\Delta t_\sigma$ \\
\midrule
1000 & 1000 & 300 & 25, 50, 100, 200 & 0.5, 2, 8 \\
\bottomrule
\end{tabular*}
\end{table}

\begin{figure}[h!]
    \begin{minipage}[c]{.49\linewidth}
      \vspace{6pt}
      \centering
      \includegraphics[scale=0.475]{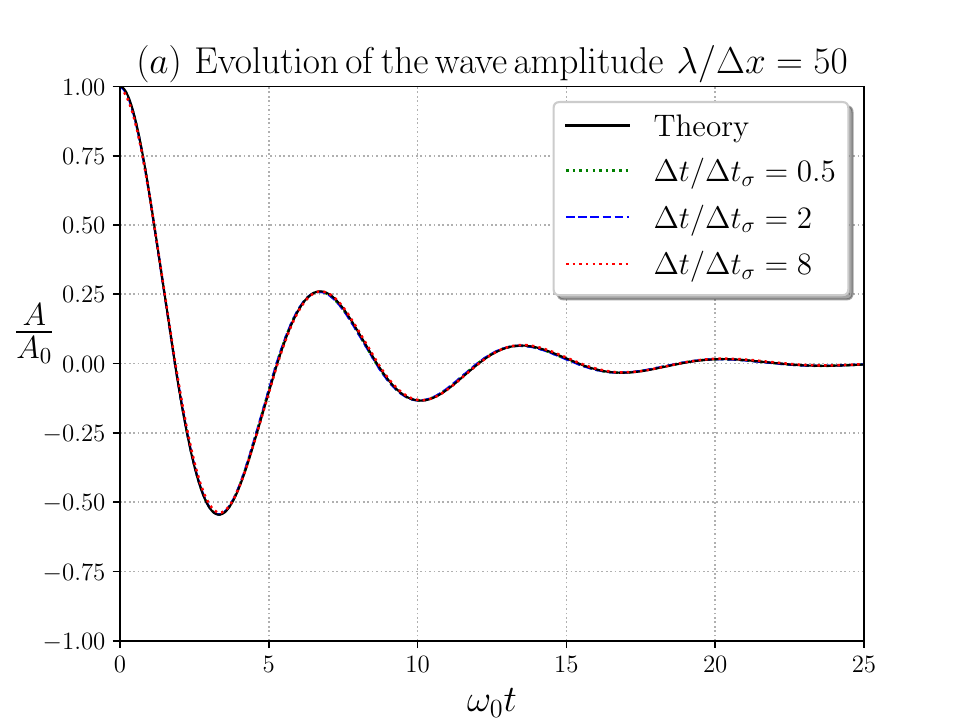}
         \centering
    \end{minipage}
    \hfill%
    \begin{minipage}[c]{.49\linewidth}
      \vspace{6pt}
      \centering
      \includegraphics[scale=0.475]{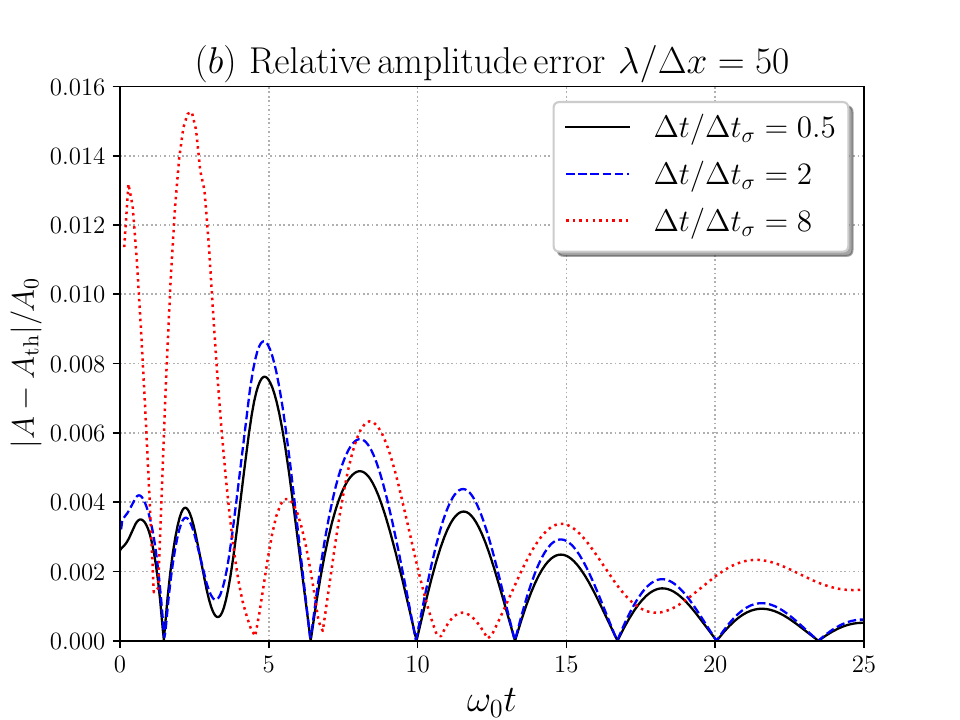}
         \centering
    \end{minipage}\newline
    
    \begin{minipage}[c]{.49\linewidth}
      \vspace{6pt}
      \centering
      \includegraphics[scale=0.475]{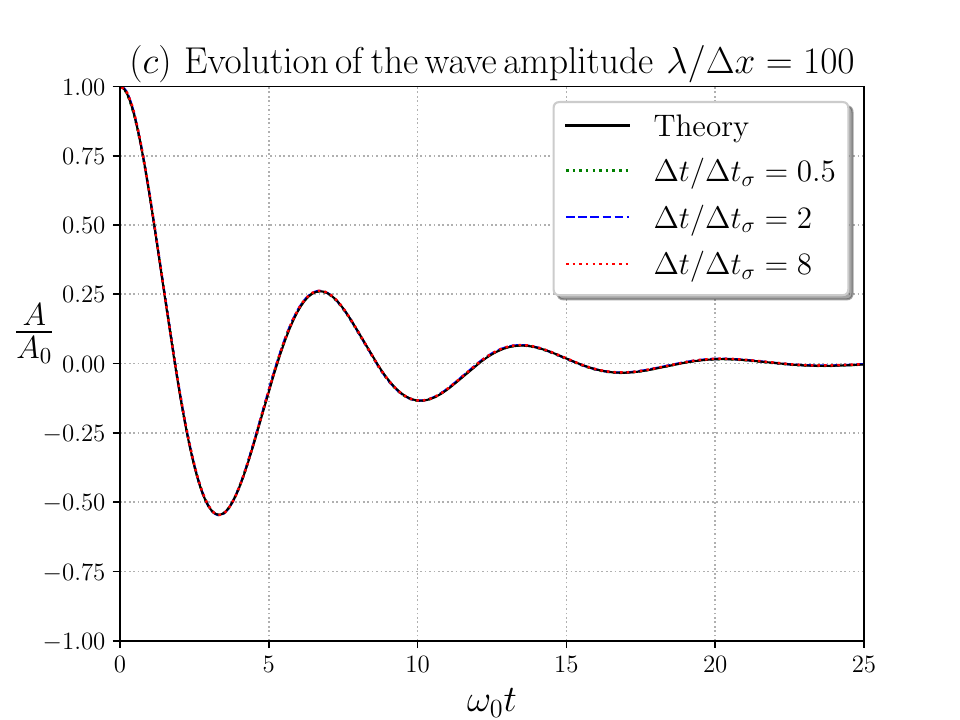}
         \centering
    \end{minipage}
    \hfill%
    \begin{minipage}[c]{.49\linewidth}
      \vspace{6pt}
      \centering
      \includegraphics[scale=0.475]{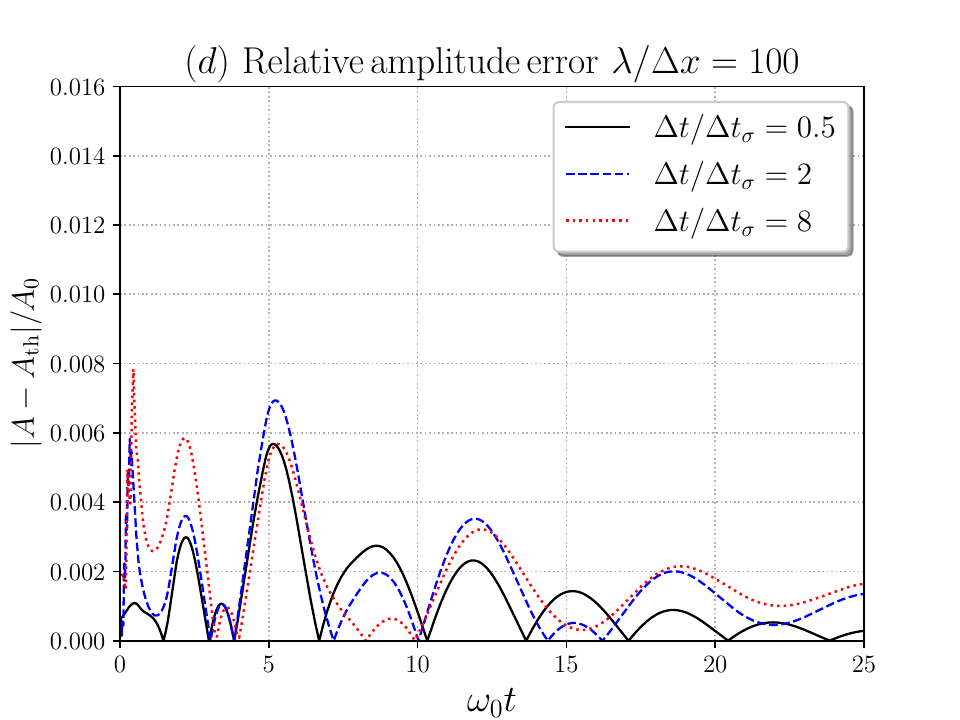}
         \centering
    \end{minipage}\newline
    
    \begin{minipage}[c]{.49\linewidth}
      \vspace{6pt}
      \centering
      \includegraphics[scale=0.475]{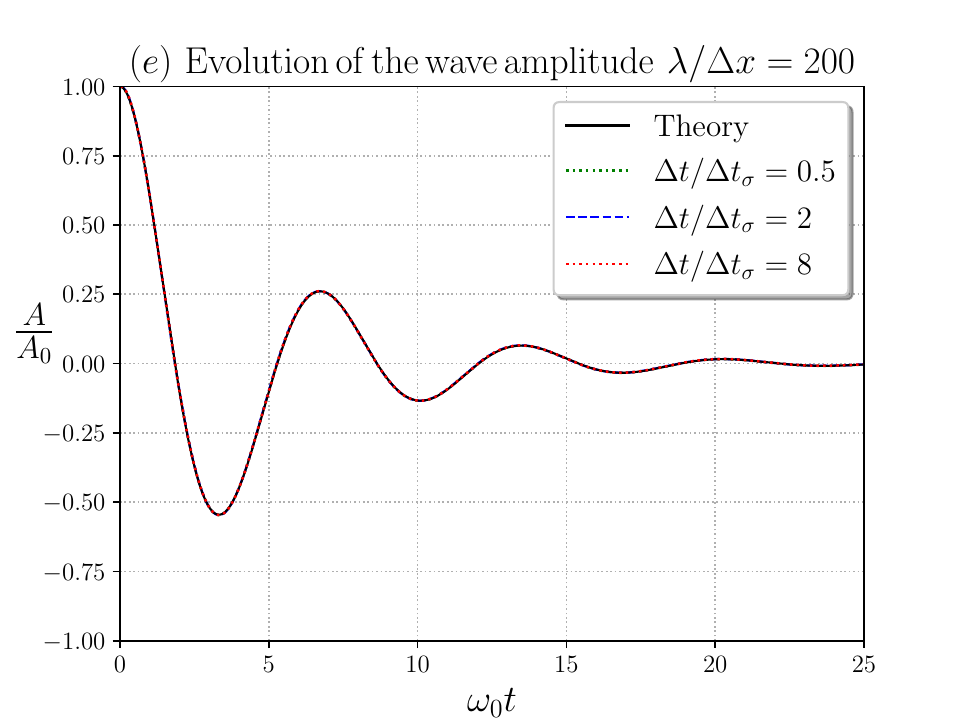}
         \centering
    \end{minipage}
    \hfill%
    \begin{minipage}[c]{.49\linewidth}
      \vspace{6pt}
      \centering
      \includegraphics[scale=0.475]{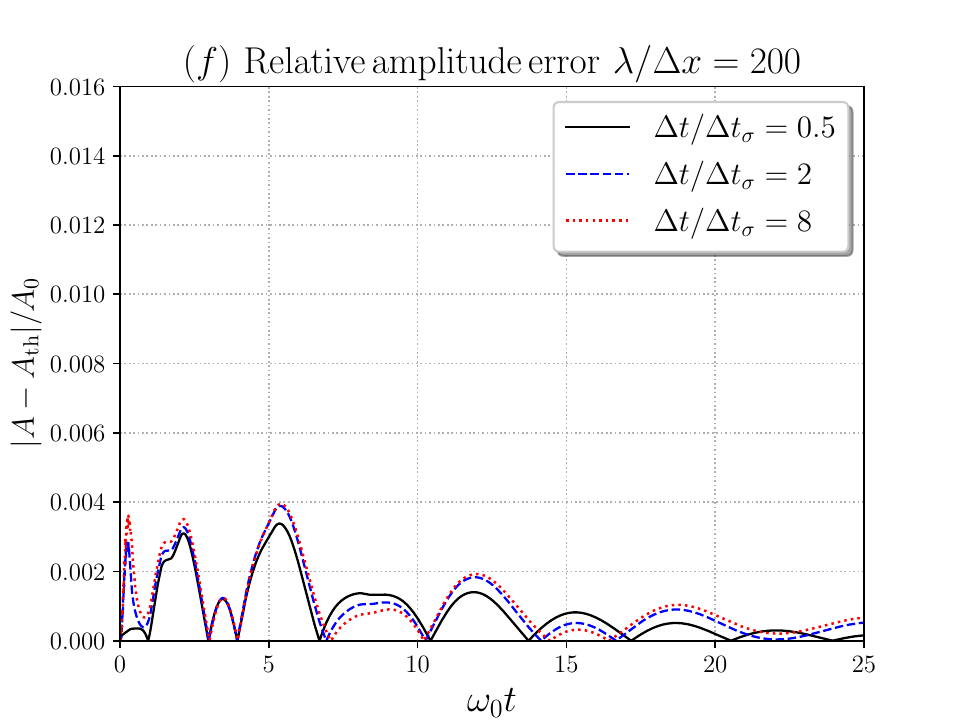}
         \centering
    \end{minipage}
    \caption{Left: Temporal evolution of the dimensionless wave amplitude $A/A_0$ over non-dimensional time $t^*=\omega_0 t$ for $\lambda/\Delta x \in\{ 50, 100, 200\}$. Right: Instantaneous relative amplitude error $|A-A_\mathrm{th}|/A_0$ for $\lambda/\Delta x\in\{  50, 100, 200\}$, where $A_\text{th}$ is the amplitude of the analytical solution of \citet{Prosperetti1981}.}
    \label{FIG:damped_wave_plot}
\end{figure}

\begin{figure}
      \begin{subfigure}{0.49\textwidth}
        \includegraphics[width=\linewidth]{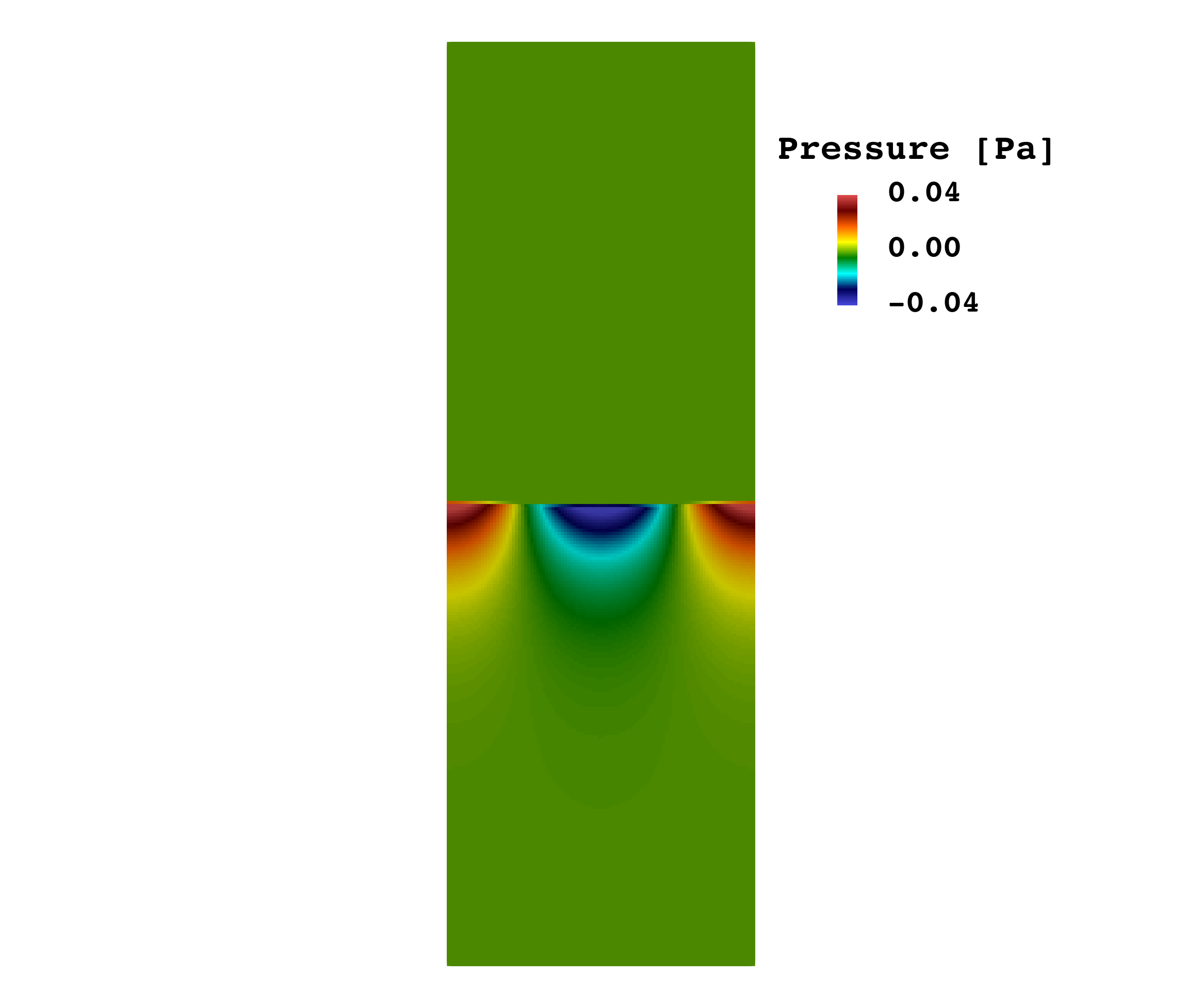}
        \caption{Pressure $p$} \label{fig:cwave_contour_p}
      \end{subfigure}%
      \hspace*{\fill}   % maximize separation between the subfigures
      \begin{subfigure}{0.49\textwidth}
        \includegraphics[width=\linewidth]{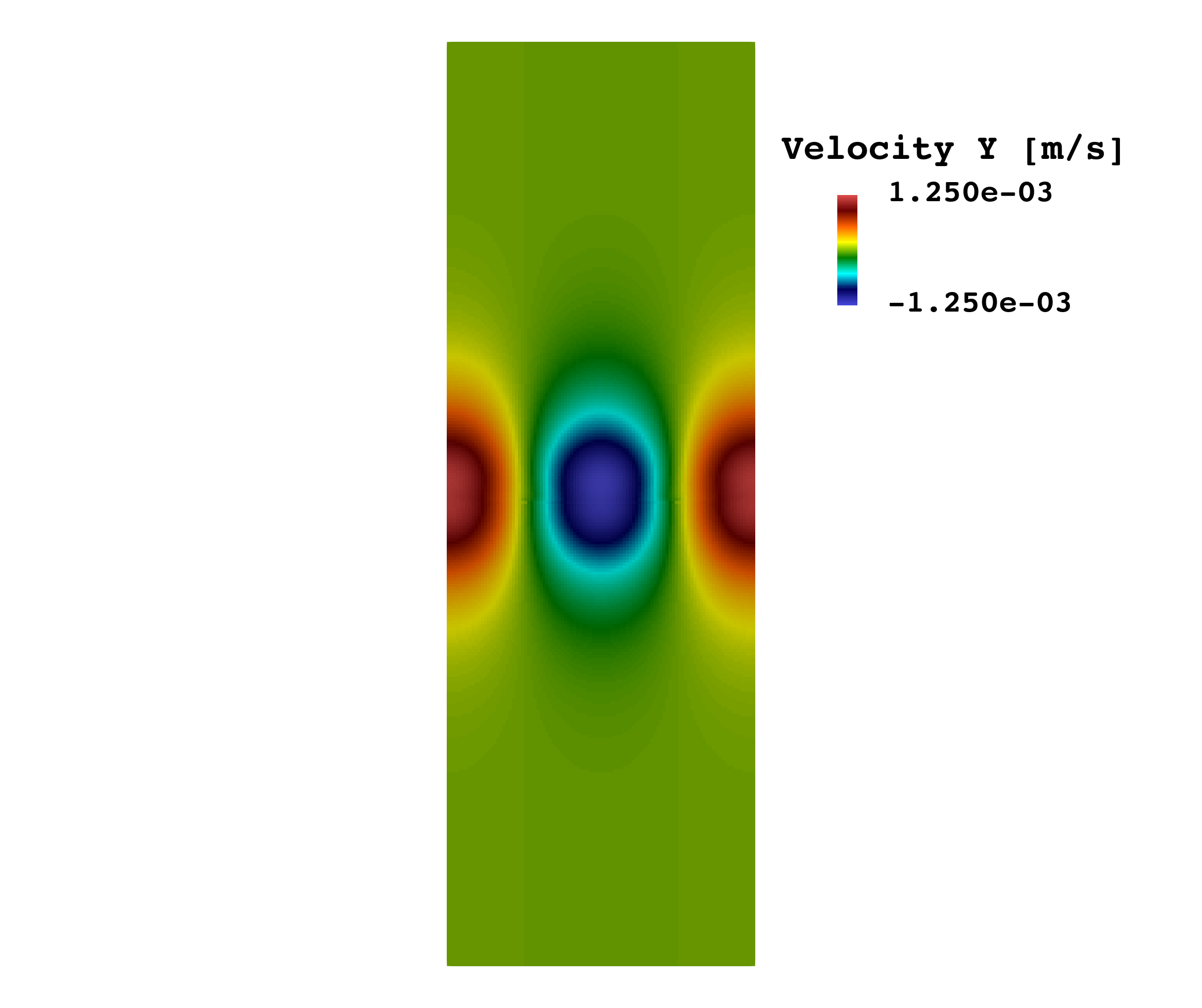}
        \caption{$y$-Velocity} \label{fig:cwave_contour_uy}
      \end{subfigure}%
      \hspace*{\fill}   % maximizeseparation between the subfigures
    \caption{\textcolor{black}{Pressure field (a) and vertical velocity field (b) at $t^*=\omega_0 t=2.5$ for $\lambda/\Delta x=100$ and $\Delta t/\Delta t_{\sigma}=8$.}} \label{fig:cwave_contours}
    \end{figure}

The temporal evolution of the wave amplitude over time is shown in Figs.~\ref{FIG:damped_wave_plot} (a), (c), and (e). The results obtained with the proposed fully-coupled algorithm are in excellent agreement with the analytical solution of \citet{Prosperetti1981}, especially for $\lambda/\Delta x \geq 50$, irrespective of the applied time step. The corresponding instantaneous relative amplitude error $|A(t)-A_\mathrm{th}(t)|/A_0$ shown in Figs.~\ref{FIG:damped_wave_plot} (b), (d), and (e) suggest that time-step independence is reached with $200$ points per wavelength. \textcolor{black}{In order to better illustrate the present test case and confirm the robustness of the solver, the pressure and vertical velocity fields are provided in Fig.~\ref{fig:cwave_contours}, at time instant $\omega_0 t=2.5$ (right before the wave reaches its minimum amplitude), for $\lambda/\Delta x=100$ and the largest applied time step, $\Delta t/\Delta t_{\sigma}=8$.}

To further quantify the differences between analytical solutions and numerical results, the $L_2$ error norm is computed relative to the initial amplitude $A_0$, as previously considered by \citet{Popinet2009}, defined as
\begin{equation}
L_2=\frac{1}{A_0}\sqrt{\frac{1}{t_\text{end}}\int_{t=0}^{t_\text{end}}\left[A(t)-A_{\mathrm{th}}(t)\right]^2 \,\text{d}t},
\label{eq:L2_damped_wave}
\end{equation}
where $t_\text{end}$ is the end time of the simulation, $A$ is the wave amplitude obtained with the proposed numerical framework and $A_{\mathrm{th}}$ is the wave amplitude obtained with the analytical solution of \citet{Prosperetti1981}. The values of the $L_2$ error norm are gathered in Table~\ref{table_2D_damped_wave_2}. The $L_2$ error norms for all cases decrease with increasing mesh resolution and decreasing time step and, in general, the $L_2$ error norms are small ($<10^{-2}$) for all considered cases, with the \textcolor{black}{expectation} of the wave on the \textcolor{black}{coarsest} mesh simulated with the largest time step. The rate of convergence is between $1/2$ and $1$ for all time steps, which is low compared to the second-order convergence observed with a unit density ratio on Cartesian meshes~\citep{Popinet2009}. However, the density ratio of $\rho_\text{A}/\rho_\text{B}=1000$ is less common in the literature and existing results with explicit surface tension~\citep{Herrmann2008a,Desjardins2008} also reported a low convergence rate (order $1$ or less) for this case. \textcolor{black}{The limiting factor regarding the order of accuracy of this test case are the interface transport and the computation of the interface curvature. The interface transport with the employed schemes is at best second order accurate, which means that the computation of the mean curvature of the interface is at best zeroth order accurate \cite{Evrard2024}. This suggests that, for a sufficiently high spatial and temporal resolution, the order of convergence of the amplitude error should be zero, meaning that the amplitude error is constant and does not reduce with further mesh refinement.}

\begin{table}[width=.9\linewidth,cols=4]
\caption{$L_2$ error norm, see Eq.~(\ref{eq:L2_damped_wave}), for each resolution and for each prescribed time step relative to the capillary time-step constraint. The corresponding order of convergence is given in parentheses.}\label{table_2D_damped_wave_2}
\begin{tabular*}{\tblwidth}{@{} LLLL@{} }
\toprule
$\lambda/\Delta x$ & $L_2\,(\Delta t/\Delta t_{\sigma} = 0.5)$ & $L_2\,(\Delta t/\Delta t_{\sigma} = 2)$ & $L_2\,(\Delta t/\Delta t_{\sigma}= 8)$ \\
\midrule
25 &  $5.02\times 10^{-3}$ & $6.00\times 10^{-3}$ & $2.96\times 10^{-2}$ \\
50 &  $2.73\times 10^{-3}$ (0.88) & $3.15\times 10^{-3}$ (0.93) & $4.67\times 10^{-3}$ (2.66) \\
100 &  $1.76\times 10^{-3}$ (0.63) & $2.29\times 10^{-3}$ (0.46) & $2.42\times 10^{-3}$ (0.95) \\
200 &  $1.17\times 10^{-3}$ (0.59) & $1.35\times 10^{-3}$ (0.76) & $1.43\times 10^{-3}$ (0.76) \\
\bottomrule
\end{tabular*}
\end{table}

\begin{figure}
\centering
		\includegraphics[scale=.65]{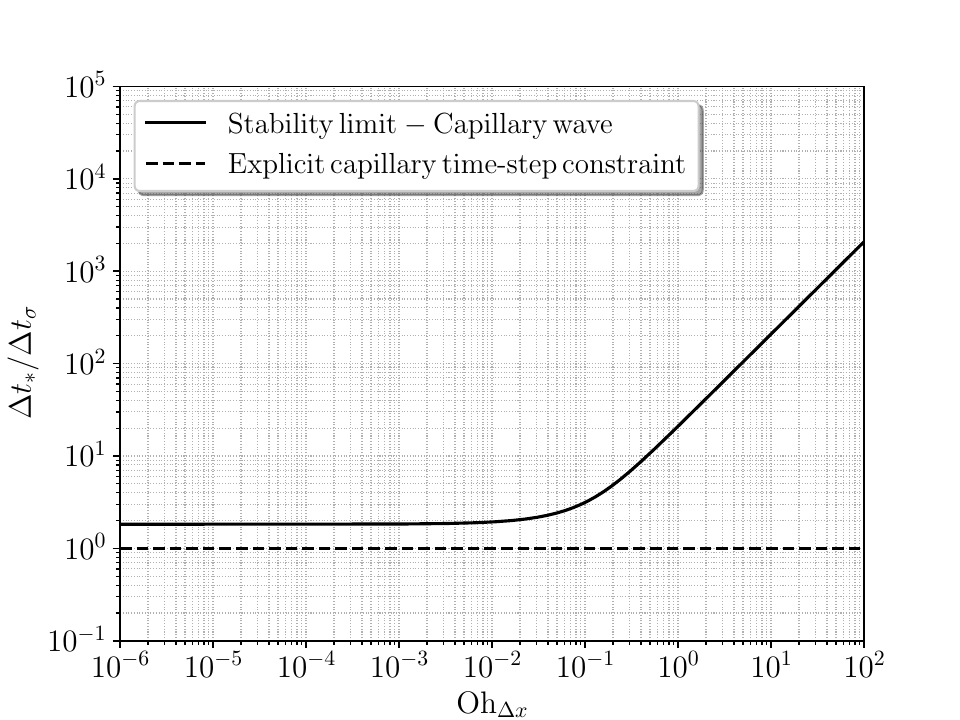}
		\caption{Approximated upper capillary stability limit as a function of the mesh Ohnesorge number, $\text{Oh}_{\Delta x}$, for the capillary wave case with density ratio $\rho_\text{A}/\rho_\text{B}=1000$.}
		\label{fig:upper_lim_capillary_wave}
\end{figure}

The upper capillary stability limit for this capillary wave case is plotted in Fig.~\ref{fig:upper_lim_capillary_wave}. Similar to the Laplace equilibrium discussed in the previous section, the stability limit is more restrictive for the density ratio of $\rho_\text{A}/\rho_\text{B}=1000$ considered here than for the density ratio $\rho_\text{A}/\rho_\text{B}=1$ considered by~\citet{Denner2022b}. Nevertheless, especially cases in which viscosity dominates, either because the dynamic viscosity is large or the capillary wave is short, benefit from a drastically increased applicable time step.

\subsection{Oscillating droplet}
\label{subsection:oscillating_droplet}

The case of an oscillating droplet, which has been previously presented in~\citep{Raessi2012a,  Vaudor2017}, considers the oscillations of a dense two-dimensional viscous droplet, in which the effects of the surface tension are examined. The initially \textcolor{black}{elliptical} droplet is situated in a unit square domain, with initial major axis $a = 0.15 \, \mathrm{m}$ and minor axis $b = 0.1 \, \mathrm{m}$. The physical properties and dimensionless parameters of the problem are summarised in Tables~\ref{table:osc_drop_phys_num_param} and~\ref{table:osc_drop_dimensionless_param}, respectively. The simulations are run until $t / \tau_\mu = 1.5$, where $\tau_\mu = \rho_\mathrm{A} D_0^2 / \mu_\mathrm{A}$ is the viscous time scale, to ensure that both spurious currents and physical interface oscillations are dissipated completely. The time step applied in these simulations is calculated as the minimum between the CFL and capillary time-step constraints: $\Delta t = \min\left(\Delta t_\mathrm{CFL}, \Sigma\Delta t_\sigma\right)$, where $\Sigma = \Delta t/\Delta t_\sigma$ is the factor by which the capillary time-step constraint $\Delta t_\sigma$ is breached. Three different time steps are considered for this test case, with $\Sigma\in\{2, 5, 10\}$, and the maximum CFL number is $0.05$.

\begin{table}[width=.9\linewidth,cols=6]
\caption{Physical parameters for the oscillating droplet test.}\label{table:osc_drop_phys_num_param}
\begin{tabular*}{\tblwidth}{@{} LLLLLL@{} }
\toprule
$\rho_\mathrm{A}$ ($\mathrm{kg.m^{-3}}$) & $\rho_\mathrm{B}$ ($\mathrm{kg.m^{-3}}$) & $\mu_\mathrm{A}$ ($\mathrm{kg.m^{-1}.s^{-1}}$) & $\mu_\mathrm{B}$ ($\mathrm{kg.m^{-1}.s^{-1}}$) & $\sigma$  ($\mathrm{N.m^{-1}}$) & $R_0=\sqrt{ab}$ ($\mathrm{m}$) \\
\midrule
$1000$ & $1$ & $7.5\times 10^{-2}$ & $7.5\times 10^{-2}$ & $0.1$ & $0.1224744871$ \\
\bottomrule
\end{tabular*}
\end{table}

\begin{table}[width=.9\linewidth,cols=5]
\caption{Dimensionless parameters for the oscillating droplet test.}\label{table:osc_drop_dimensionless_param}
\begin{tabular*}{\tblwidth}{@{} LLLL@{} }
\toprule
$\rho_\mathrm{A} / \rho_\mathrm{B}$ & $\mu_\mathrm{A} / \mu_\mathrm{B}$ & $\mathrm{La} = \rho_\mathrm{A}\sigma R_0/\mu_\mathrm{A}^2$ & $R_0 / \Delta x$ \\
\midrule
$1000$ & $1$ & $2177$ & $7.8$  \\
\bottomrule
\end{tabular*}
\end{table}

\begin{figure}
\centering
		\includegraphics[scale=0.225]{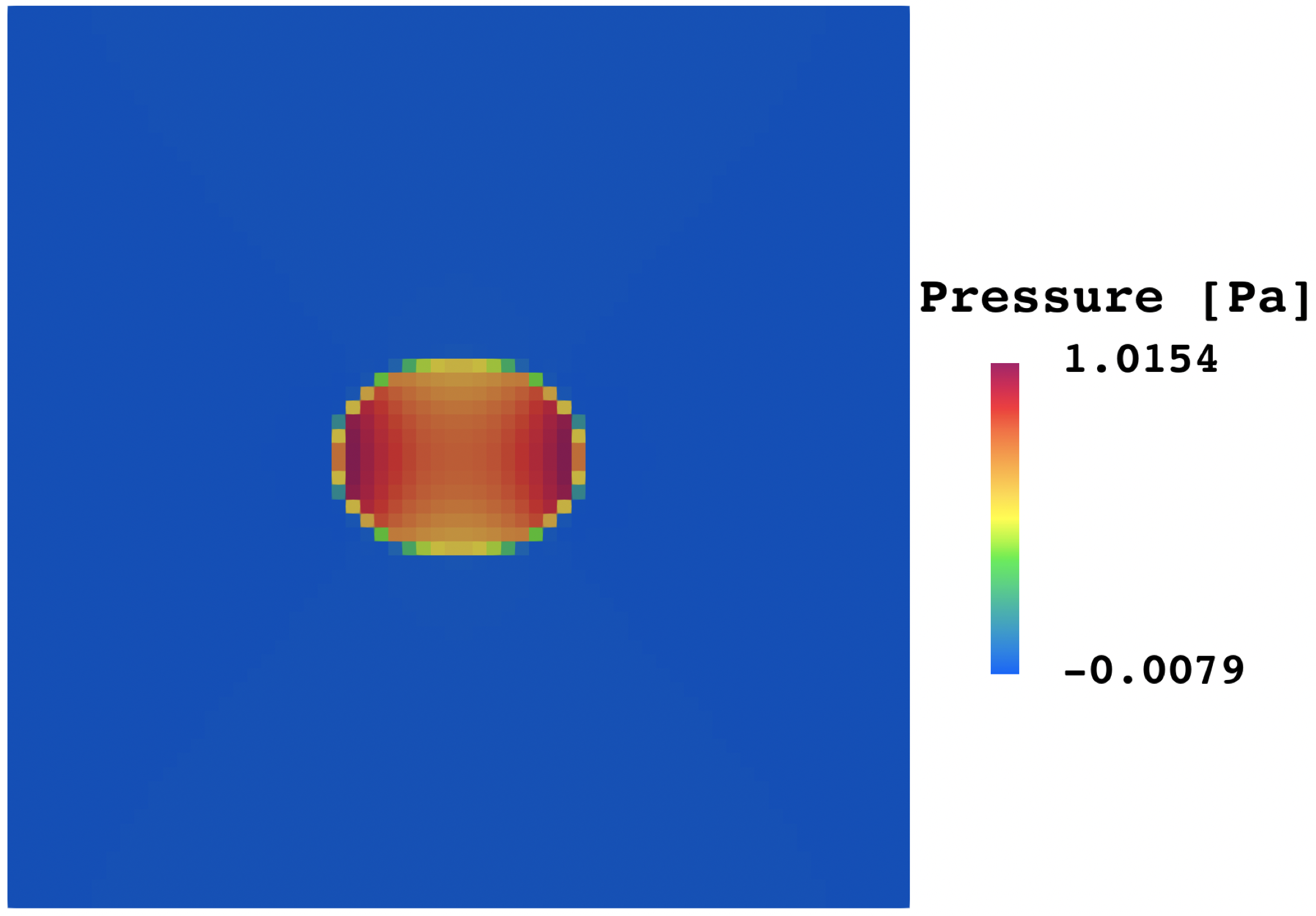}
		\caption{\textcolor{black}{Pressure field of the 2D oscillating droplet at $t^*=t/\tau_\mu=0.015$ with $\Sigma = 10$.}}
		\label{fig:osc_drop_pressure_contour}
\end{figure}

Over time, the oscillations of the droplet induced by surface tension are damped by viscous stresses. Using the theoretical work of~\citet{Rush2000} and~\citet{Lamb1932}, the evolution of the dimensionless total energy (i.e.~potential and kinetic) of the 2D droplet for oscillation mode $n$ is given as
\begin{equation}
    E^*(t) = \frac{E(t)}{E_0} = \exp\left(-\frac{2n(n-1)\mu_\mathrm{A} t}{\sqrt{\rho_\mathrm{A}\sigma R_0}}\right) = \exp\left(-\frac{2n(n-1) t}{\sqrt{\mathrm{La}}}\right).    \label{eq:Lamb_total_energy}
\end{equation}
The initial total energy, $E_0$, which corresponds to the initial potential energy associated with surface tension, is calculated as $E_0 = \sigma (\mathcal{C}_\mathrm{ellipse} - \mathcal{C}_\mathrm{circle})$, where $\mathcal{C}_\mathrm{ellipse}$ is the circumference of the ellipse and $\mathcal{C}_\mathrm{circle}$ is the circumference of the circle with the same area. Following the work of~\citet{Lamb1932}, the analytical expression for the oscillation frequency of the $n$-th mode is given as 
\begin{equation}
    \omega_n = \sqrt{\frac{n(n-1)(n+1)\sigma}{(\rho_\mathrm{A} + \rho_\mathrm{B})R_0^3}},
    \label{eq:Lamb_osc_freq}
\end{equation}
where $R_0= \sqrt{ab}$ is the equilibrium radius of the droplet. As previously considered by \citet{Raessi2012a} and \citet{Vaudor2017}, the dominant second mode ($n=2$) is studied here. 
\textcolor{black}{To illustrate the test case, the pressure field during the second oscillation of the droplet, at $t^*=t/\tau_\mu=0.015$, with the largest prescribed time step, $\Sigma = 10$, is shown in Fig.~\ref{fig:osc_drop_pressure_contour}.}

\begin{figure}
\begin{minipage}[c]{.48\linewidth}
    \vspace{6pt}
    \centering
    \includegraphics[scale=0.55]{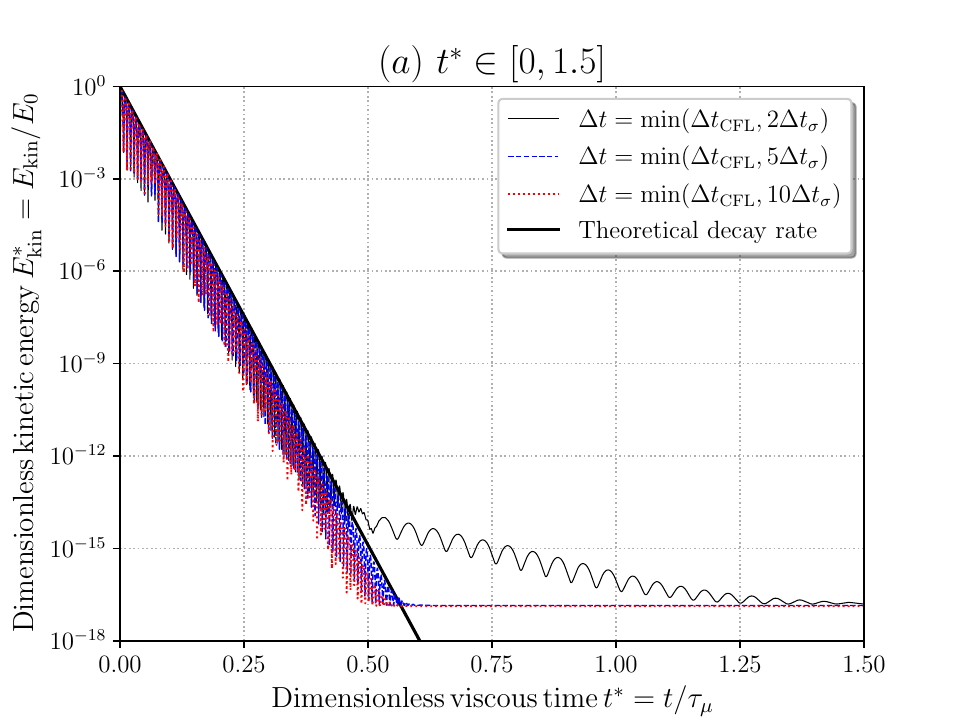}
       \centering
    \end{minipage}
    \hfill%
    \begin{minipage}[c]{.48\linewidth}
    \vspace{6pt}
      \centering
    \includegraphics[scale=0.55]{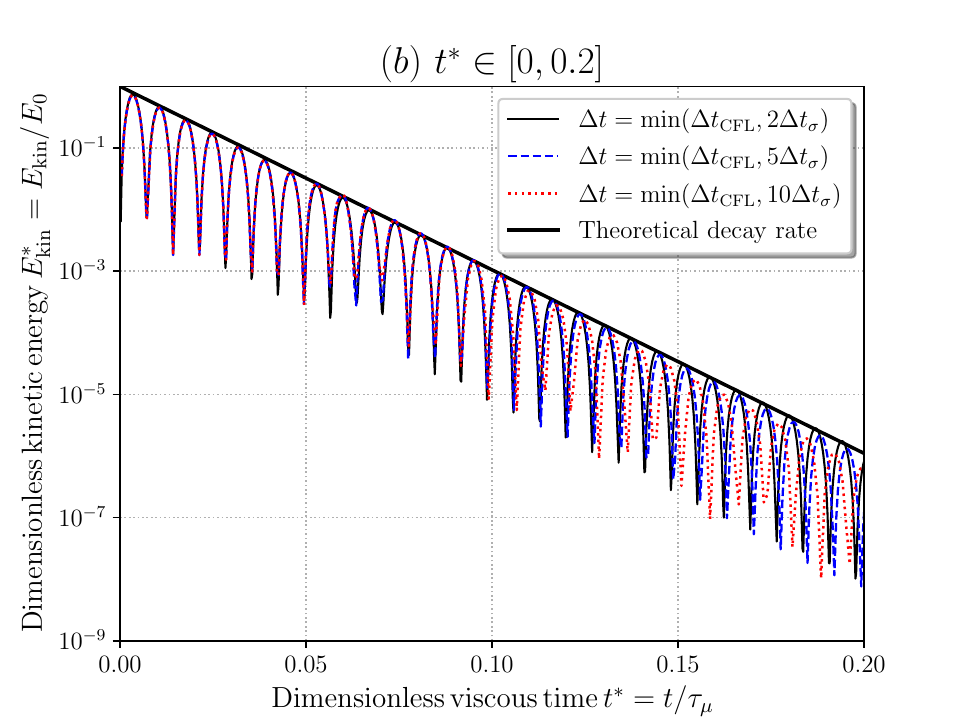}
       \centering
    \end{minipage}
    \caption{Temporal evolution of the dimensionless kinetic energy of the 2D oscillating droplet over time for (a) the entire simulation and (b) a zoomed interval.}
    \label{fig:kin_energy_osc_drop}
\end{figure}

\begin{figure}
\centering
		\includegraphics[scale=.65]{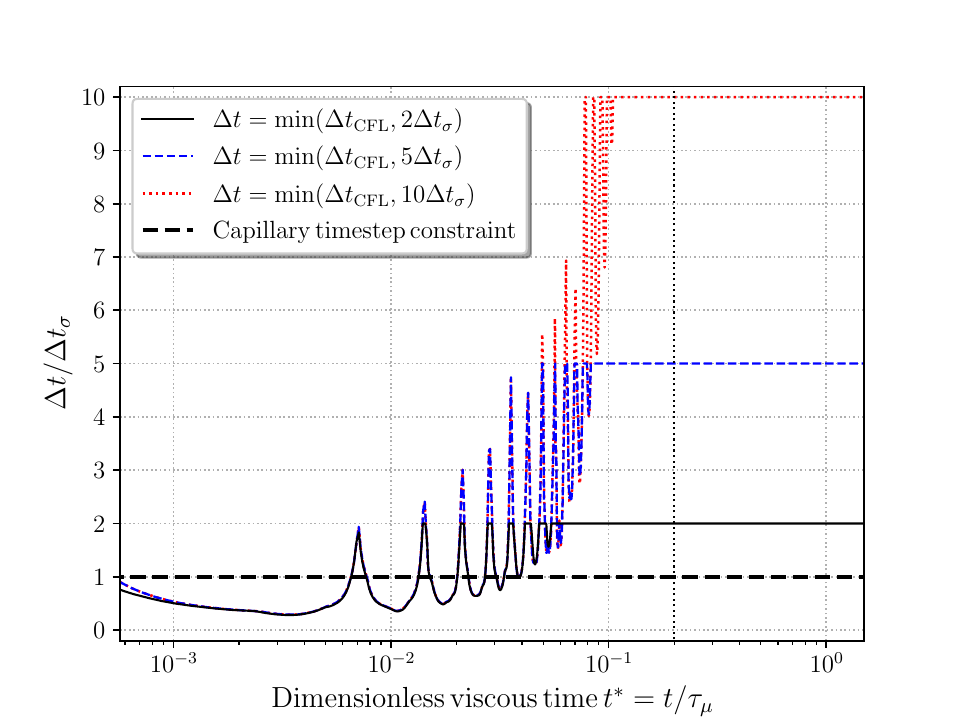}
		\caption{Temporal evolution of the applied time step, normalised by the capillary time-step constraint, for each run ($\Sigma = {2, 5, 10}$) of the 2D oscillating droplet. The dotted vertical black line corresponds to $t^* = 0.2$, which is the upper bound of the zoomed interval in Fig.~\ref{fig:kin_energy_osc_drop} (b).}
		\label{fig:osc_drop_dt_all}
\end{figure}

The temporal evolution of the dimensionless kinetic energy of the droplet $E^*_\mathrm{kin} = E_\mathrm{kin} / E_0$ for all simulations is shown in Fig.~\ref{fig:kin_energy_osc_drop} (a). A zoomed view is provided in Fig.~\ref{fig:kin_energy_osc_drop} (b) over the range $t^*\in[0, 0.2]$, corresponding to the interval in which all but one millionth of the initial droplet energy $E_0$ dissipates. The temporal evolution of the effective time step of the simulations is displayed in Fig.~\ref{fig:osc_drop_dt_all}, which confirms the representative choice of the reduced interval $t^*\in[0, 0.2]$, used for the calculations of errors following below. Despite a strong CFL constraint, the largest part of the simulations is conducted with a time step larger than the capillary time-step constraint, as observed in Fig.~\ref{fig:osc_drop_dt_all}. At the end of all simulations the kinetic energy is, as expected, fully dissipated, see Fig.~\ref{fig:kin_energy_osc_drop} (a). The theoretical decay rate is accurately reproduced for all time steps in the range $t^* =t/\tau_\mu < 0.1$, as seen in Fig.~\ref{fig:kin_energy_osc_drop} (b). Fig.~\ref{fig:osc_drop_dt_all} indicates that $t^*=0.1$ corresponds approximately to the time after which the time step reaches its maximum value $\Delta t = \Sigma \Delta t_\sigma$. A more significant difference in the quality of the results (oscillation frequency and kinetic energy decay) depending on the time step is observed for $t^* > 0.1$.

\begin{figure}
\begin{minipage}[c]{.48\linewidth}
    \vspace{6pt}
    \centering
    \includegraphics[scale=0.55]{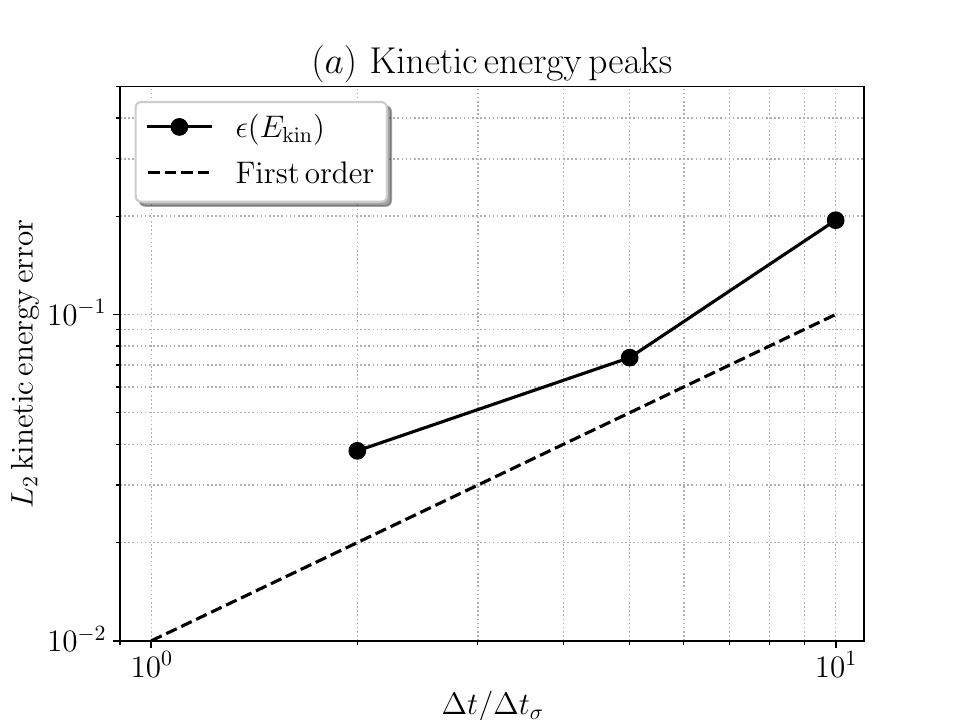}
       \centering
    \end{minipage}
    \hfill%
    \begin{minipage}[c]{.48\linewidth}
    \vspace{6pt}
      \centering
    \includegraphics[scale=0.55]{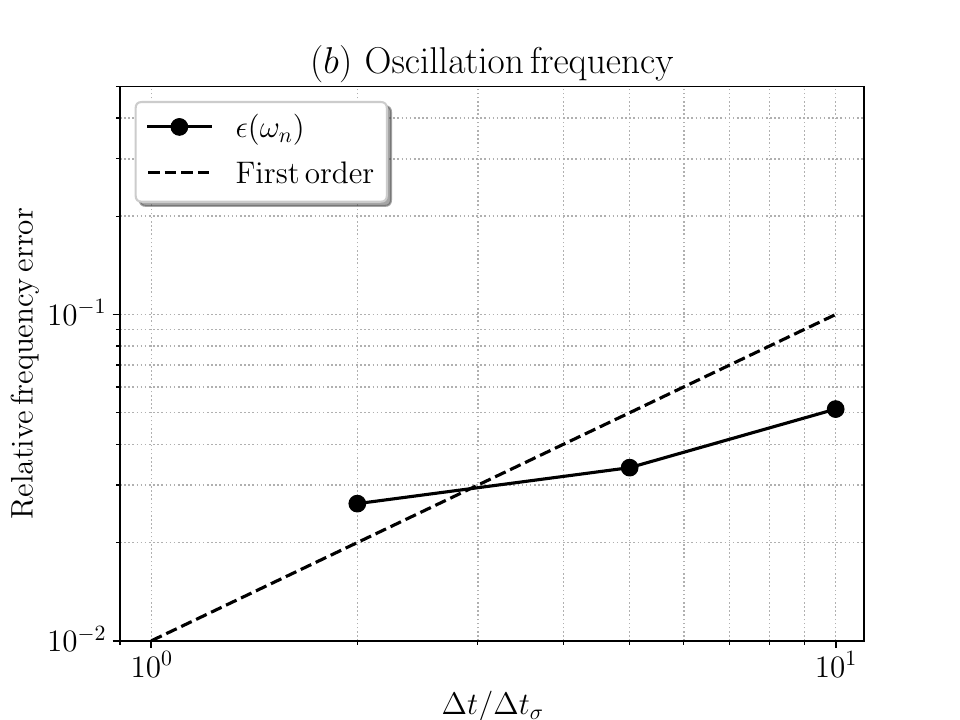}
       \centering
    \end{minipage}
    \caption{Temporal convergence of the oscillating droplet test: (a) the error in kinetic energy, defined by Eq.~\eqref{eq:error_osc_drop_kin_energy}, and (b) the error in oscillation frequency, defined by Eq.~\eqref{eq:error_osc_drop_freq}.}
    \label{fig:kin_energy_osc_freq_cv}
\end{figure}

The errors of the oscillation frequency $\omega_n$ and kinetic energy $E_\mathrm{kin}$ are computed over the range $t^*\in [0, 0.2]$ as
\begin{equation}
    \epsilon(\omega_n) = \frac{\omega_{n,\text{calc}} - \omega_{n,\text{Lamb}}}{\omega_{n,\text{Lamb}}} = \frac{\left(2\pi/\mathcal{T}_{n,\text{calc}}\right) - \omega_{n,\text{Lamb}}}{\omega_{n,\text{Lamb}}},
    \label{eq:error_osc_drop_freq}
\end{equation}
and
\begin{equation}
    \epsilon(E_\mathrm{kin}) = L_2(E_\mathrm{kin}) = \sqrt{\frac{1}{N_{\mathrm{peaks}}}\sum_{p=1}^{N_{\mathrm{peaks}}}\left(\frac{E^*_{\mathrm{kin, calc},p}-E^*_{\mathrm{kin, th},p}}{E^*_{\mathrm{kin, th},p}}\right)^2}
    \label{eq:error_osc_drop_kin_energy}
\end{equation}
with $\mathcal{T}_{n,\text{calc}}$ the average period calculated over the first 14 oscillations (corresponding to the range $t^*\in [0, 0.2]$) for each simulation, $N_\text{peaks} = 28$ (corresponding to $t^*\in [0, 0.2]$), and $E^*_{\mathrm{kin, calc}}$ and $E^*_{\mathrm{kin, th}}$ are the dimensionless energy obtained from the simulations and from theory, respectively. The error of the final radius is also computed at the end of each simulation, $t_\text{end} = 1.5 t^*$ as
\begin{equation}
    \epsilon(R) = \frac{\left|R_{\text{calc}}(t_\text{end}) - R_0\right|}{R_0} = \frac{\left|\sqrt{a_\text{calc}(t_\text{end})\,b_\text{calc}(t_\text{end})} - \sqrt{ab}\right|}{\sqrt{ab}},
    \label{eq:error_radius}
\end{equation}
where $a_{\text{calc}}$ and $b_{\text{calc}}$ are calculated by summing up the colour function row-wise and column-wise on the employed Cartesian mesh, respectively. The error values and associated convergence rate are gathered in Table~\ref{table:osc_drop_results} for all considered time steps. The kinetic-energy and oscillation-frequency errors are also plotted in Fig.~\ref{fig:kin_energy_osc_freq_cv}. The geometric error in final radius is small for all considered time steps, which shows that a circle faithfully representing the expected theoretical circle of radius $R_0$ is obtained at the end of the simulations. The very low convergence rate of the radius further indicates that the final radius of the simulation does not depend strongly on the applied time step. Fig.~\ref{fig:kin_energy_osc_freq_cv} shows that the oscillation frequency is less affected than the kinetic energy by an increase in time step. Both errors are, nonetheless, small for $\Sigma=2$ and $\Sigma=5$, while $\Sigma=10$ is not sufficient to faithfully capture the physics, as confirmed by Fig.~\ref{fig:kin_energy_osc_drop} (a). It should be noted that the errors in oscillation frequency for $\Sigma=2$ and $\Sigma=5$ (approx.~$3\%$) are smaller than those obtained by~\citep{Vaudor2017} with an explicit surface tension treatment at the same resolution, but over a shorter time interval (approx.~$4.5\%$), demonstrating the good temporal accuracy of the proposed fully-coupled algorithm.
\textcolor{black}{It should be noted that the formal order of convergence of a finite-difference scheme can, in general, only be achieved if all relevant physical process are adequately resolved. Here, however, the surface-tension-driven interface motion is not adequately resolved in time, since the capillary time-step constraint is breached. The formal second-order accuracy of the employed temporal discretisation can, thus, not be expected to be achieved.}

\begin{table}[width=.9\linewidth,cols=8]
\caption{Errors $\epsilon$ and corresponding convergence rate of the simulation of the oscillating droplet with different time steps. A negative convergence rate indicates a decrease in accuracy as the time step increases.}\label{table:osc_drop_results}
\begin{tabular*}{\tblwidth}{@{} LLLLLLLL@{} }
\toprule
$\Sigma=\Delta t/ \Delta t_\sigma$ & $\epsilon(R)$ & $\mathcal{O}(R)$ & $\epsilon(E_\mathrm{kin})$ & $\mathcal{O}(E_\mathrm{kin})$ & $\mathcal{T}_{n,\text{calc}}$ $[\mathrm{s}]$ & $\epsilon(\omega_n)$ & $\mathcal{O}(\omega_n)$ \\
\midrule
%$1$ & $?$ & $?$ & $?$ & $-$ & $?$ & $?$ & $-$\\
$2$ & $9.92\times 10^{-4}$ & $-$ & $0.038$ & $-$ & $11.297$ & $+0.0263$ & $-$\\
$5$ & $8.68\times 10^{-4}$ & $0.15$ & $0.074$ & $-0.73$ & $11.386$ & $+0.0339$ & $-0.28$\\
$10$ & $1.30\times 10^{-3}$ & $-0.58$ & $0.195$ & $-1.40$ & $11.595$ & $+0.0513$ & $-0.60$\\
\bottomrule
\end{tabular*}
\end{table}

The performance of the simulations is quantified by the \ac{wct} and the \ac{rct}, calculated as~\citep{Janodet2022}
\begin{equation}
    \mathrm{RCT} = \frac{\mathrm{WCT}\times N_\text{cores}}{N_{\Delta t} \times N_\text{cells}},
    \label{eq:RCT}
\end{equation}
where $N_\text{cores}$ is the number of processing cores, $N_{\Delta t}$ refers to the number of time steps, $N_\text{cells}$ is the number of mesh cells, and presented in Table~\ref{table:osc_drop_perfos}.
The WCT of the cases with $\Sigma=2$ and $\Sigma=5$ show that a speed up of factor $1.9$ in total simulation time is obtained when multiplying the maximum allowed time step by $2.5$. This speed up is however not maintained when increasing the maximum time step further to $\Sigma=10$, which is explained by the on average slower convergence of the nonlinear procedure within each time step for $\Sigma=10$. This is confirmed by the $\mathrm{RCT}$ values, showing a substantial increase in average reduced time spent for each time step for $\Sigma=10$. Hence, there exists an optimal case-dependent value for $\Sigma$ that maximises the performance, while still providing a sufficient resolution of the dominant transient flow features. For the present test case, $\Sigma=5$ seems to be a good compromise.

\begin{table}[width=.9\linewidth,cols=6]
\caption{Computational performance of the simulations of the oscillating droplet test for different time steps, quantified by the wall clock time (WCT) and the reduced computational time (RCT), see Eq.~\eqref{eq:RCT}.}\label{table:osc_drop_perfos}
\begin{tabular*}{\tblwidth}{@{} LLLLLL@{} }
\toprule
$\Sigma$ & $\mathrm{WCT}$ $[\mathrm{s}]$ & $N_\text{cores}$ & $N_{\Delta t}$ & $N_\text{cells}$ & $\mathrm{RCT}$ $[\mathrm{ms}]$ \\
\midrule
$2$ & $3370$ & $24$ & $8180$ & $4096$ & $2.41$ \\
$5$ & $1790$ & $24$ & $3677$ & $4096$ & $2.85$ \\
$10$ & $1410$ & $24$ & $2207$ & $4096$ & $3.74$ \\
\bottomrule
\end{tabular*}
\end{table}

\subsection{3D Rayleigh-Plateau instability}
\label{subsection:rayleigh_plateau}

The Rayleigh-Plateau instability is studied to demonstrate the capability of the proposed numerical framework with the implicit surface tension framework to predict surface-tension-driven flows in three dimensions that feature a significant deformation of the liquid-gas interface. To this end, a cylindrical filament of radius $r_0$ is initialised in a cubic domain with a small longitudinal perturbation imposed at the interface, 
\begin{equation}
    r(z) = r_0 + \epsilon r_0\cos(kz) = r_0 + A_0\cos(kz),
    \label{eq:RP_1}
\end{equation}
where $\epsilon = 0.03$, $r_0 = 0.2$, and $k = \pi$. Following \citet{Popinet2009}, in order to compare the results to linear theory, the velocity field is initialised as $\mathbf{u}=\boldsymbol{\nabla}\phi$, with the potential
\begin{equation}
    \phi = -\frac{A_0 c}{ikJ_1(ikr_0)}J_0(ikr)\cos(kz),
    \label{eq:RP_2}
\end{equation}
where $c$ is the inviscid growth rate obtained from linear stability analysis \citep{Rayleigh1892}
\begin{equation}
\left(\frac{c}{c_0}\right)^2 = \frac{I_1(\xi)}{I_0(\xi)}\xi(1-\xi^2),
\label{eq:RP_3}
\end{equation}
with $c_0^2=\sigma/(\rho_\mathrm{A} r_0^3)$ and $\xi=kr_0$ the dimensionless wavenumber. To include viscous effects, Weber reformulated the growth-rate equation as~\citep{Weber1931, Sterling1975}:
\begin{equation}
c^2 + 3\nu_\mathrm{A} k^2 c = \frac{1}{2}c_0^2\xi^2(1-\xi^2) ,
\label{eq:RP_Weber_growth_rate}
\end{equation}
which allows to validate cases with low to moderate Laplace numbers. Following previous studies \citep{Dai2005,Popinet2009}, the Laplace number considered here is $\mathrm{La}=238$. 

\begin{table}[width=.9\linewidth,cols=5]
    \caption{Dimensionless parameters for the Rayleigh-Plateau instability.}\label{table:RP_dimensionless_param}
    \begin{tabular*}{\tblwidth}{@{} LLLLL@{} }
    \toprule
    $\rho_\mathrm{A} / \rho_\mathrm{B}$ & $\mu_\mathrm{A} / \mu_\mathrm{B}$ & $\mathrm{La} = \rho_\mathrm{A}\sigma r_0/\mu_\mathrm{A}^2$ & $\xi=kr_0$ & $r_0 / \Delta x$ \\
    \midrule
    $832$ & $56$ & $238$ & $0.628$ & $20$  \\
    \bottomrule
    \end{tabular*}
    \end{table}
    
    \begin{figure}
      \begin{subfigure}{0.31\textwidth}
        \includegraphics[width=\linewidth]{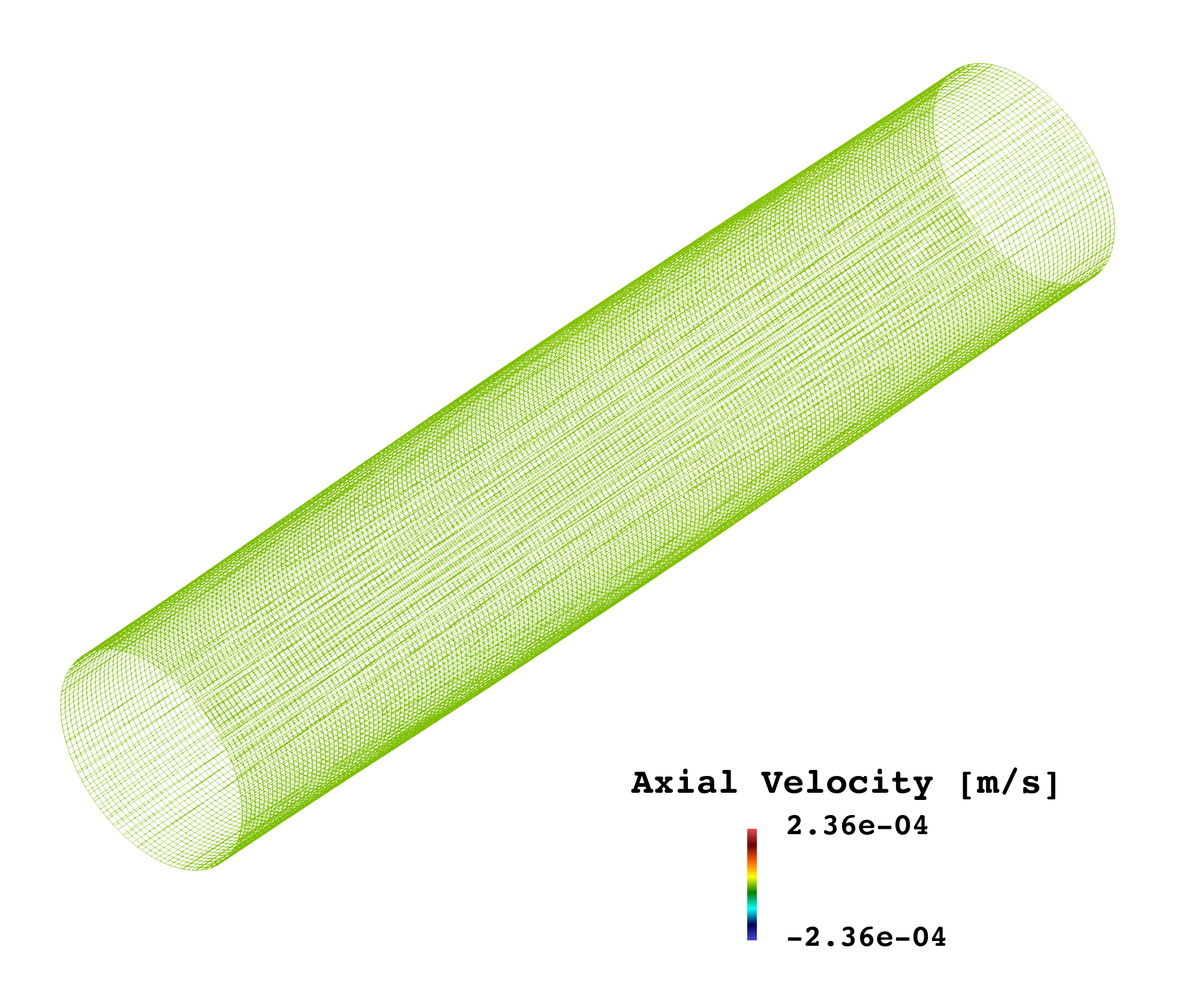}
        \caption{$t/T_\sigma=0$} \label{fig:RP_contour_1}
      \end{subfigure}%
      \hspace*{\fill}  
      \begin{subfigure}{0.31\textwidth}
        \includegraphics[width=\linewidth]{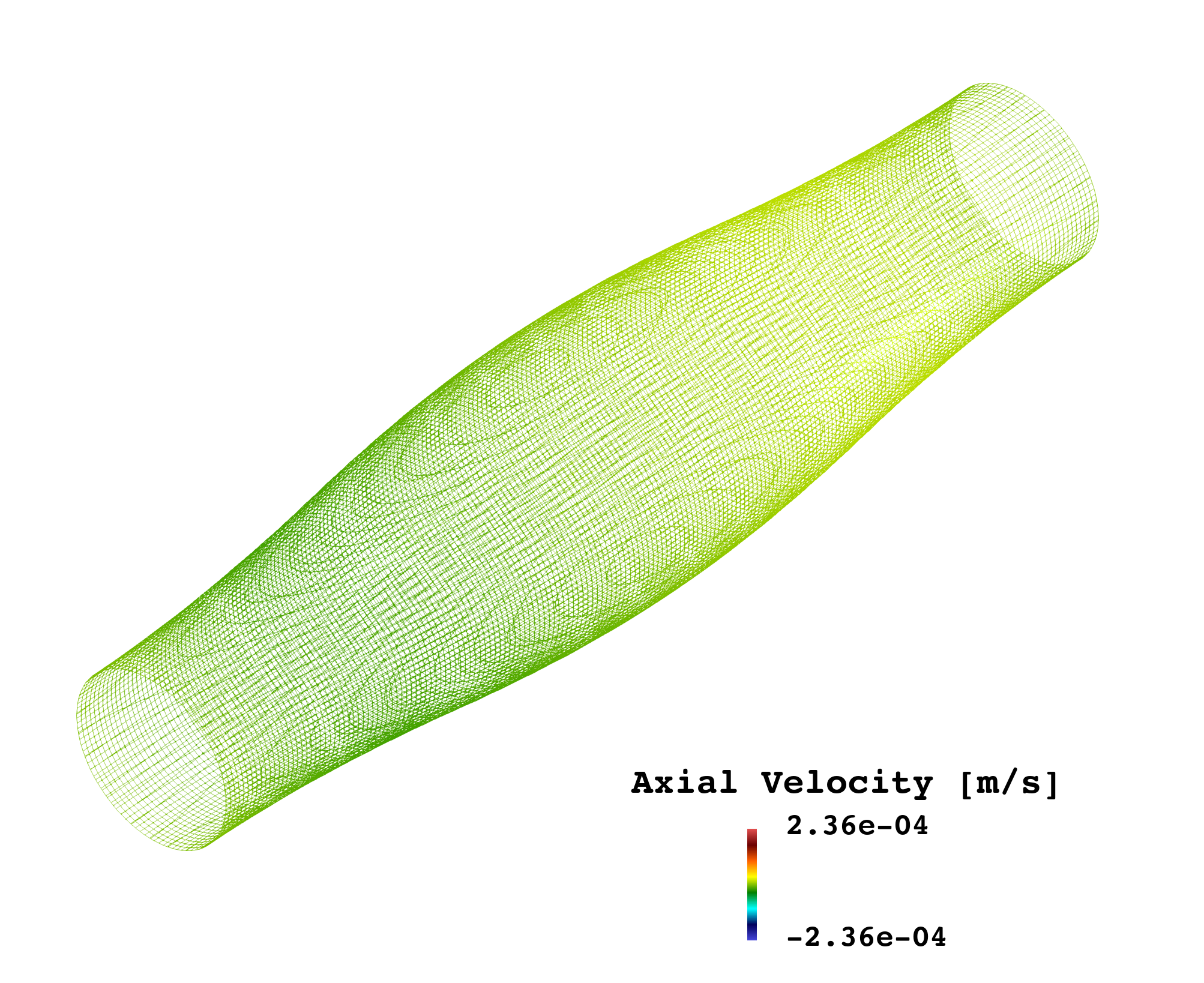}
        \caption{$t/T_\sigma=6$} \label{fig:RP_contour_2}
      \end{subfigure}%
      \hspace*{\fill}  
      \begin{subfigure}{0.31\textwidth}
        \includegraphics[width=\linewidth]{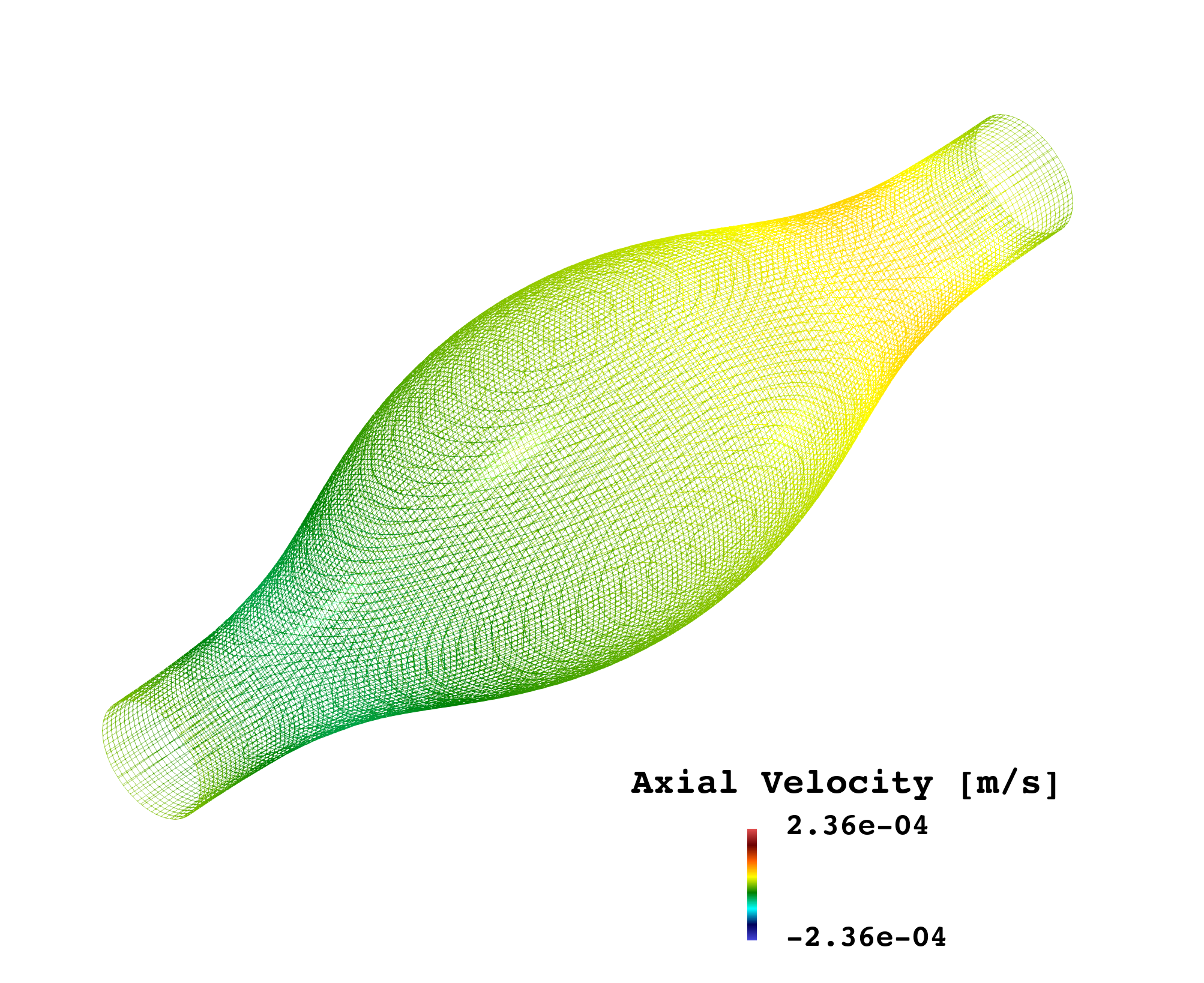}
        \caption{$t/T_\sigma=9$} \label{fig:RP_contour_3}
      \end{subfigure}\\
      \begin{subfigure}{0.31\textwidth}
        \includegraphics[width=\linewidth]{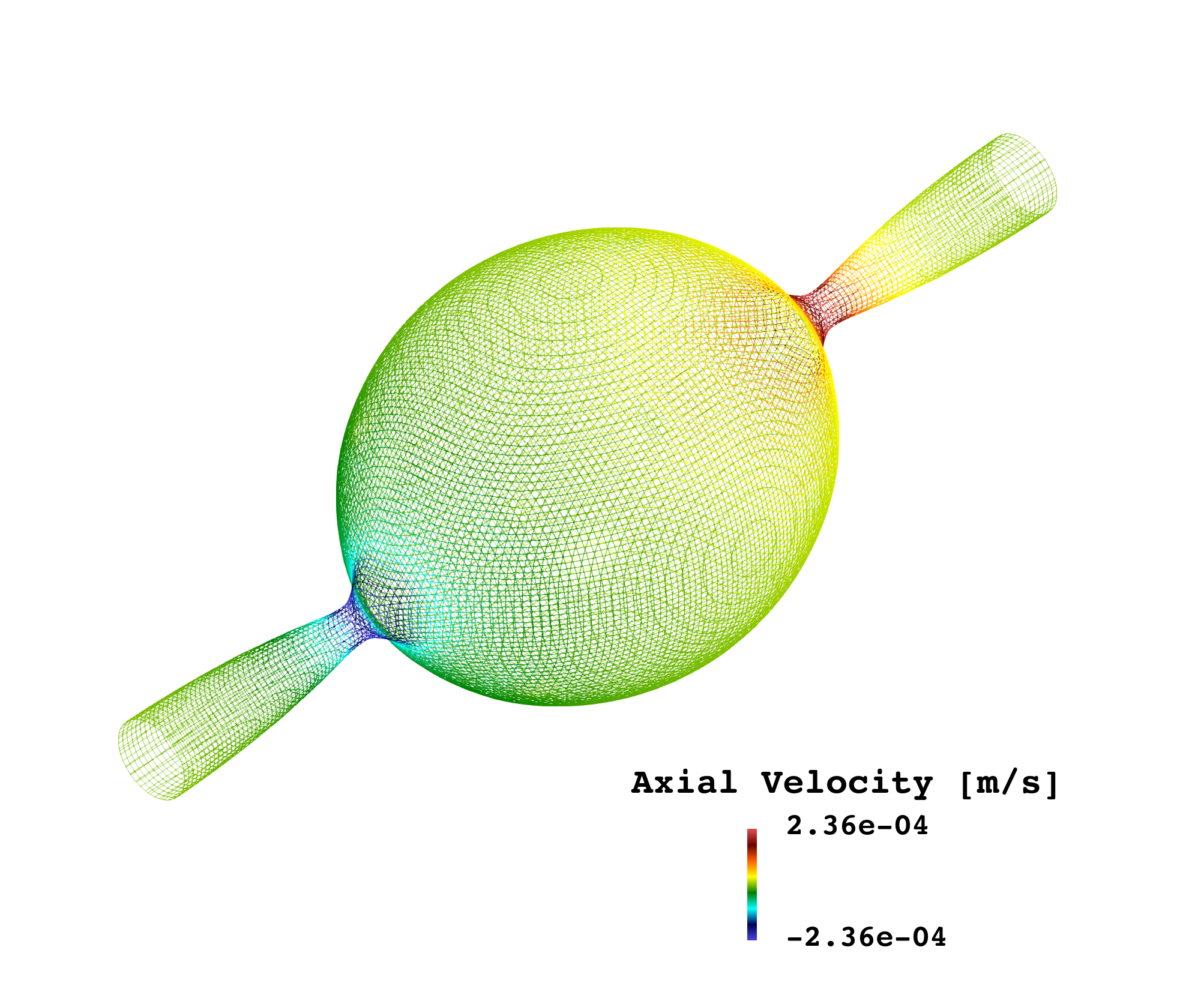}
        \caption{$t/T_\sigma=10.8$} \label{fig:RP_contour_4}
      \end{subfigure}%
      \hspace*{\fill}  
      \begin{subfigure}{0.31\textwidth}
        \includegraphics[width=\linewidth]{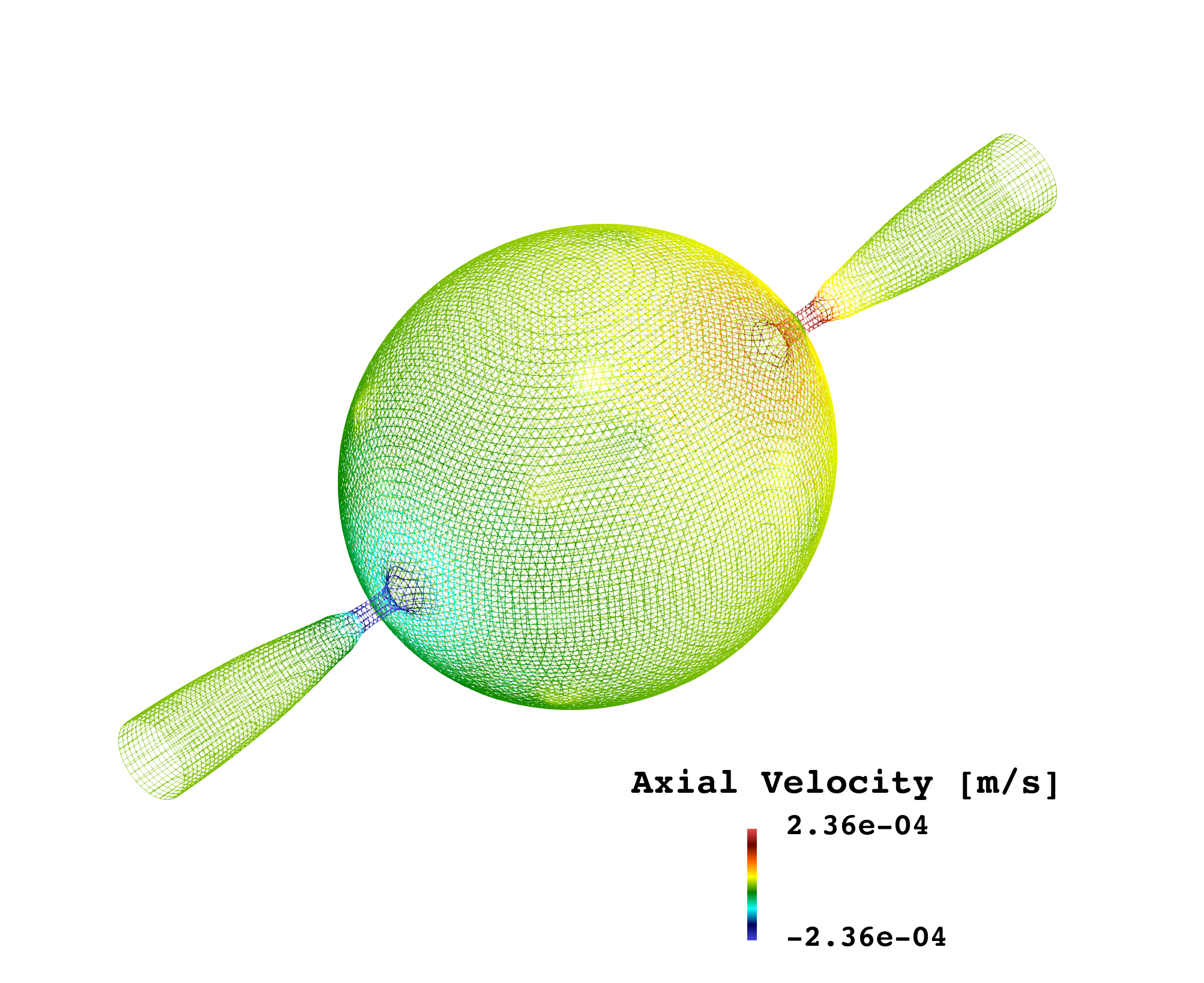}
        \caption{$t/T_\sigma=11$} \label{fig:RP_contour_5}
      \end{subfigure}%
      \hspace*{\fill}  
      \begin{subfigure}{0.31\textwidth}
        \includegraphics[width=\linewidth]{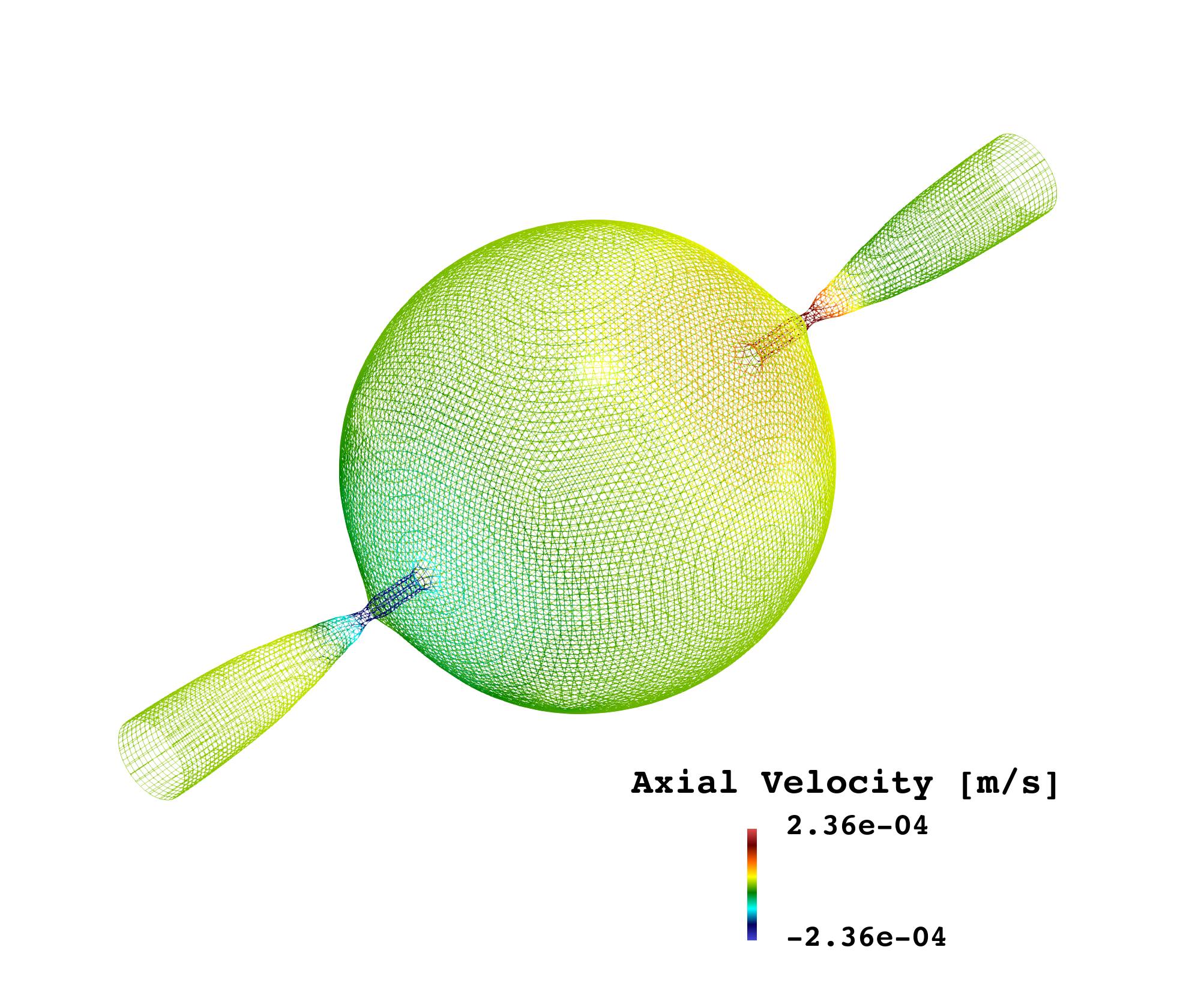}
        \caption{$t/T_\sigma=11.2$} \label{fig:RP_contour_6}
      \end{subfigure}
    \caption{Interface contours of the Rayleigh-Plateau instability of a cylindrical filament for different time instances for $\Sigma = 2$, coloured by the axial velocity.} \label{fig:RP_contours}
    \end{figure}

    \begin{figure}
        \begin{minipage}[c]{.48\linewidth}
            \vspace{6pt}
            \centering
            \includegraphics[scale=0.55]{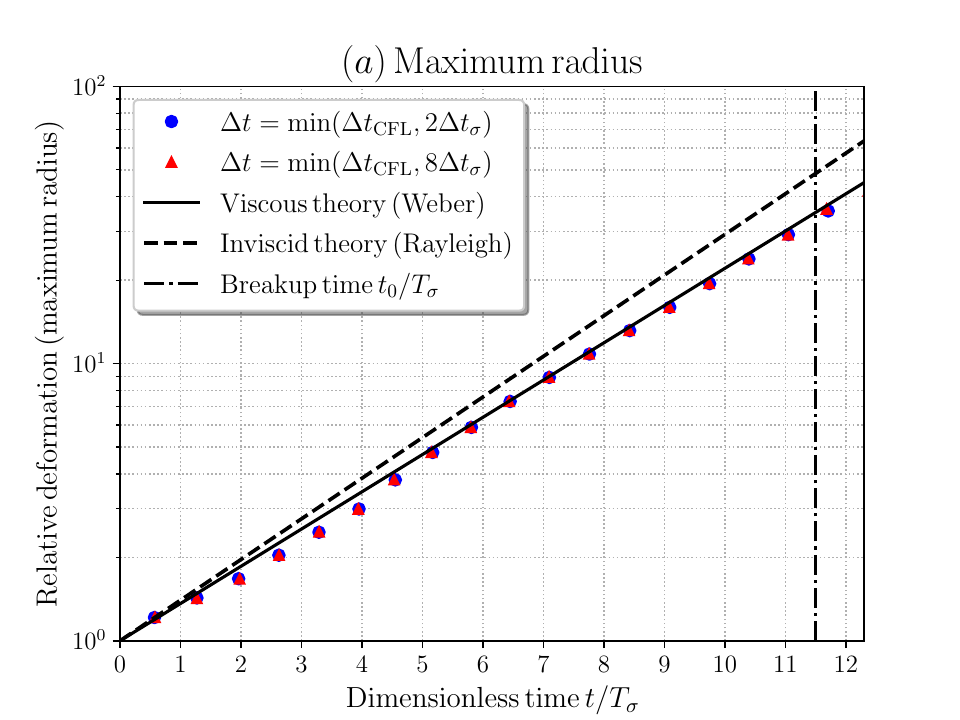}
               \centering
            \end{minipage}
            \hfill%
            \begin{minipage}[c]{.48\linewidth}
            \vspace{6pt}
              \centering
            \includegraphics[scale=0.55]{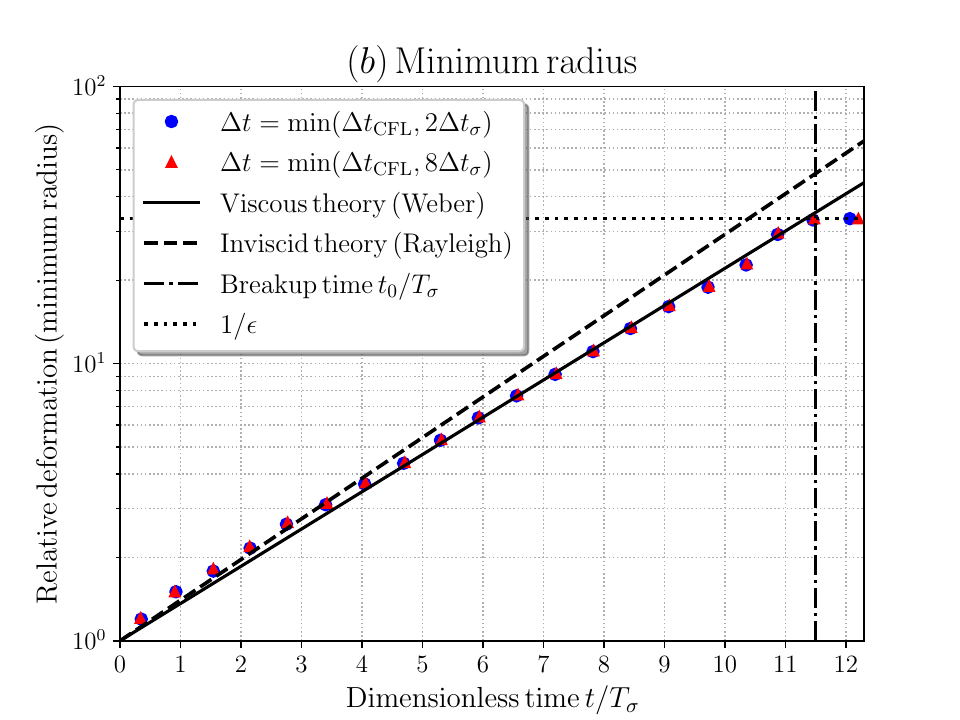}
               \centering
            \end{minipage}
            \caption{Relative deformation of the liquid filament over dimensionless time $t/T_\sigma$ based on (a) the maximum radius $(r_\mathrm{max}-r_0)/(\epsilon r_0)$ and (b) the minimum radius $(r_0-r_\mathrm{min})/(\epsilon r_0)$, compared to Weber's viscous linear theory, defined in Eq.~(\ref{eq:RP_Weber_growth_rate}). The dotted line in (b) indicates the theoretical relative deformation from breakup (i.e. when $r_\mathrm{min}=0$).}
            \label{fig:RP_Weber_scaling}
        \end{figure}

        \begin{figure}
            \centering
                    \includegraphics[scale=.65]{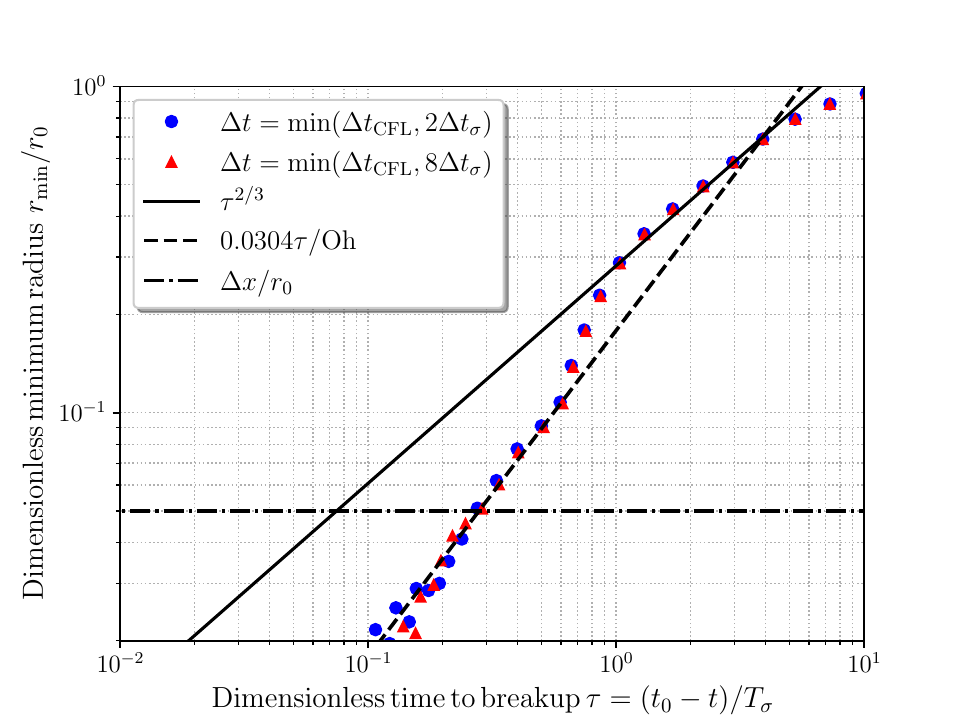}
                    \caption{Evolution of the dimensionless minimum radius $r_\mathrm{min}/r_0$ with respect to the dimensionless time to breakup $\tau = (t_0 - t)/T_\sigma$. The dash-dotted line $\Delta x/r_0$ indicates the spatial resolution limit.}
                    \label{fig:RP_scalings_final}
            \end{figure}
        
        \begin{figure}
        \centering
                \includegraphics[scale=.65]{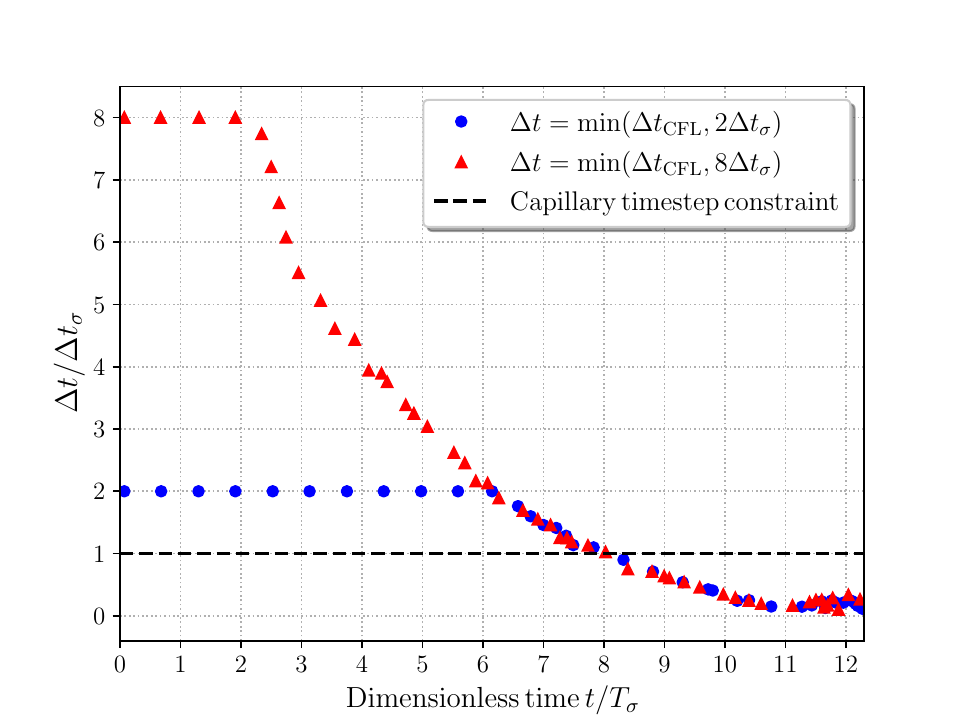}
                \caption{Temporal evolution of the applied time step, normalised by the capillary time-step constraint, for both simulations, $\Sigma \in \{2, 8\}$, of the Rayleigh-Plateau instability.}
                \label{fig:RP_dt_over_time}
        \end{figure}

A quarter of the filament is simulated in a $[0; 0.5] \times [0; 0.5] \times [0; 1]$ domain with symmetric boundary conditions, using an equidistant Cartesian mesh with $50 \times 50 \times 100$ cells, which corresponds to 20 cells per initial filament radius. The same dimensional parameters as previously considered by~\citet{Popinet2009} are adopted for this test case, summarised in Table~\ref{table:RP_dimensionless_param}. The characteristic time scale $T_\sigma = 1/c_0$ of the Rayleigh-Plateau instability is defined by the growth rate $c_0$. The time step applied in the simulations is calculated as the minimum between the CFL and capillary time-step constraints: $\Delta t = \min\left(\Delta t_\mathrm{CFL}, \Sigma\Delta t_\sigma\right)$, where $\Sigma = \Delta t/\Delta t_\sigma$ is the factor by which the capillary time-step constraint $\Delta t_\sigma$ is breached. Two different time steps are considered for this test case, with $\Sigma\in\{2, 8\}$, and the maximum CFL number is $0.04$. The interface contours illustrating the development of the Rayleigh-Plateau instability and the induced interface deformation are shown for $\Sigma=2$ in Fig.~\ref{fig:RP_contours}.

Following~\citet{Popinet2009}, the relative deformation of the interface, calculated based on both the maximum and minimum radii, is compared to Weber's viscous linear theory~\citep{Weber1931,  Sterling1975} in Fig.~\ref{fig:RP_Weber_scaling}. The results are in excellent agreement for both considered time steps, demonstrating the ability of the proposed coupled solver to faithfully reproduce capillary effects in three dimensions with significant interface deformation. The non-dimensional breakup time predicted with both time steps is $t_0/T_\sigma=11.5$, which is slightly earlier than a breakup time of $t_0/T_\sigma=12.1$ reported by~\citet{Popinet2009} with explicit surface tension. We attribute this difference to the coarser mesh resolution used here compared to the work of \citet{Popinet2009}, which means the neck of the filament reaches a radius that cannot be resolved  and, consequently, succumbs to numerical breakup at an earlier time $t_0/T_\sigma$. Following \citet{Denner2022}, the evolution of the dimensionless minimum radius $r_\mathrm{min}/r_0$ as a function of the dimensionless time to pinch-off $\tau=(t_0-t)/T_\sigma$ is plotted in Fig.~\ref{fig:RP_scalings_final}. Early on in the evolution of the Rayleigh-Plateau instability, the minimum radius follows the inertial scaling $r_\mathrm{min}\sim (t_0 - t)^{2/3}$ \citep{Castrejon-Pita2015}, whereas shortly before pinch-off the minimum radius faithfully reproduces the theoretical inertial-viscous solution defined as~\citep{Eggers1993} 
\begin{equation}
\frac{r_\mathrm{min}}{r_0} = \frac{0.0304\tau}{\mathrm{Oh}},
    \label{eq:inertial_viscous_theory_RP}
\end{equation}
where $\mathrm{Oh}=1/\sqrt{\mathrm{La}}$ is the Ohnesorge number. 
The temporal evolution of the applied time step is shown in Fig.~\ref{fig:RP_dt_over_time}. The calculations are conducted with a time step larger than the capillary time-step up to $t/T_\sigma=8$, at which point the CFL constraint becomes the dominant time-step constraint. Hence, the majority of the simulations is carried out with a time step larger than the capillary time-step constraint.

\section{Conclusions}
\label{section:conclusion}

The performance of computer simulations of surface-tension-driven phenomena is often limited, in many cases severely, by the capillary time-step constraint \citep{Brackbill1992,Popinet2018}. Mitigating or eliminating this constraint, therefore, promises significant performance gains, for applications including microfluidic interfacial flows, quasi-stationary evaporation, {\color{black} dynamic contact line problems}, and spray atomization. 

In this article, a fully-coupled pressure-based algorithm for interfacial flows with implicit surface tension treatment has been 
derived, implemented and validated.
In the present work, the THINC/QQ scheme \citep{Xie2017,  Chen2022b} has been used to discretise the advection term of the colour function transport equation, instead of the CICSAM scheme~\citep{Ubbink1999} considered previously in similar algorithms \citep{Denner2014a,Denner2022b}, with the aim of alleviating the stringent CFL number constraint of $\text{CFL} \lesssim 0.01$ associated with the CICSAM scheme.
However, even with the THINC/QQ scheme, which is able to retain a sharp interface for CFL numbers of $\mathcal{O}(0.1)$, stable results could only be obtained for a maximum CFL number of $0.05$ in the current study.

In conclusion, the proposed algorithm allows simulating realistic gas-liquid flows with time steps larger than the capillary time-step constraint, as long as other time-step constraints are satisfied. Nevertheless, areas of further improvement remain to maximise the potential of the proposed numerical approach. In order to handle  complex topology changes, the implicit \textcolor{black}{height-function} method requires an implicit alternative to increase its robustness in the case of under-resolved interfaces. In addition, further improvements with respect to the robustness of the interface advection scheme should enable simulations with larger CFL numbers, bearing the potential to greatly improve the performance of the proposed algorithm.

\section*{Acknowledgements}
This research was funded by the Deutsche Forschungsgemeinschaft (DFG, German Research Foundation), grant numbers  452036112, 452916560, and 458610925. Fruitful discussions with Aman Jain and Fabien Evrard are gratefully acknowledged. Data supporting this publication can be obtained from \url{https://zenodo.org/records/13215768} under a Creative Commons Attribution license.

\end{document}